\documentclass[aps,twocolumn,showpacs]{revtex4}
\usepackage{amsmath}
\usepackage{epsfig}
\usepackage{multirow}
\usepackage{appendix}
\usepackage{amssymb}

\begin{document}
	
\title{Investigating excited $\Omega_c$ states from pentaquark perspective}
\author{Ye Yan$^1$}\email{221001005@njnu.edu.cn}
\author{Xiaohuang Hu$^2$}\email{xiaohuanghu@foxmail.com}
\author{Hongxia Huang$^1$}\email{hxhuang@njnu.edu.cn(Corresponding author)}
\author{Jialun Ping$^1$}\email{jlping@njnu.edu.cn(Corresponding author)}
\affiliation{$^1$Department of Physics, Nanjing Normal University, Nanjing, Jiangsu 210097, P. R. China}
\affiliation{$^2$Changzhou Institute of Industry Technology, Changzhou, Jiangsu 213164, P. R. China}

\begin{abstract}
Inspired by the recent observation of new $\Omega_c^0$ states by the LHCb Collaboration, we explore the excited $\Omega_{c}$ states from the pentaquark perspective
in the quark delocalization color screening model.
Our results indicate that the $\Omega_c(3185)$ can be well interpreted as a molecular $\Xi D$ predominated resonance state with $J^P=1/2^-$.
The $\Omega_c(3120)$ can also be interpreted as a molecular $\Xi_c^* \bar{K}$ state with $J^P=3/2^-$ and a new molecular state $\Xi^*_c \bar{K}^*$ with $J^P=5/2^-$ and a mass of 3527 MeV is predicted, which is worth searching in the future. Other reported $\Omega_c$ states cannot be well described in the framework of pentaquark systems in present work. The three-quark excited state, or the unquenched picture may be a good explanation, which is worth further exploration.
\end{abstract}

	
\maketitle

\setcounter{totalnumber}{5}
	
\section{\label{sec:introduction}Introduction}

In the last few decades, significant experimental progress has been made in the sector of heavy baryons.
Many heavy baryons have been reported, such as $\Lambda_{c}$ and $\Sigma_c$ family~\cite{Knapp:1976qw,ARGUS:1993vtm,CLEO:1994oxm,ARGUS:1997snv,E687:1993bax,CLEO:2000mbh,LHCb:2017jym,BaBar:2006itc,Belle:2021qip,Belle:2014fde,Ammosov:1993pi,CLEO:1996czm,Belle:2004zjl,Belle:2022hnm}, $\Xi_c$ family~\cite{LHCb:2020gge,ALICE:2021bli,Belle:2016lhy,CLEO:1998wvk,Belle:2013htj,Belle:2020ozq,LHCb:2020iby,Belle:2020tom,Belle:2016tai} and $\Omega_c$ family~\cite{Belle:2021gtf,LHCb:2017uwr,BaBar:2006pve,LHCb:2021ptx,Belle:2017ext}.
These observations have stimulated extensive interest in understanding the structures of these baryons.
For one thing, verifying these heavy baryons could deepen our understanding of the non-perturbative behavior of quantum chromodynamics (QCD)~\cite{Chen:2016qju,Swanson:2006st,Voloshin:2007dx,Chen:2016heh,Huang:2023jec,Esposito:2016noz,Lebed:2016hpi,Guo:2017jvc}.
For another thing, with the appearance of heavy baryons that are difficult to be interpreted simply as traditional three-quark baryons, the study of multi-quark explanation has become a non-negligible subject.

Among them, the excited $\Omega_c$ baryons were considerably enriched by the LHCb Collaboration in 2017~\cite{LHCb:2017uwr}.
Five narrow $\Omega_c^0$ states were observed in the $\Xi_c^+ K^-$ invariant mass spectrum, which are $\Omega_c^0(3000)$, $\Omega_c^0(3050)$, $\Omega_c^0(3065)$, $\Omega_c^0(3090)$ and $\Omega_c^0(3120)$.
The narrow width of these states, along with their unknown quantum numbers and structures, has attracted broad interest in theoretical work.
A classical way to describe these $\Omega_c$ baryons is considering that they are conventional three-quark excitations, and another way is treating them as multi-quark states.

On the basis of three-quark configuration, $\Omega_c$ states have been studied in the framework of the Lattice QCD~\cite{Padmanath:2017lng,Bahtiyar:2020uuj}, the QCD sum rules~\cite{Agaev:2017jyt,Wang:2017zjw,Aliev:2017led,Agaev:2017lip,Wang:2017xam}, the light cone QCD sum rules~\cite{Chen:2017sci,Aliev:2018uby}, the heavy hadron chiral perturbation theory~\cite{Cheng:2017ove}, the Regge phenomenology~\cite{Oudichhya:2021kop,Oudichhya:2023awb}, the chiral quark model~\cite{Yang:2017rpg}, the constituent quark model~\cite{Wang:2017hej,Wang:2017kfr,Yao:2018jmc}, the quark-diquark model~\cite{Wang:2017vnc,Ali:2017wsf}, the quark pair creation model~\cite{Chen:2017gnu}, the $^3P_0$ model~\cite{Zhao:2017fov,Garcia-Tecocoatzi:2022zrf}, the chiral quark-soliton model~\cite{Kim:2017jpx,Kim:2017khv}, the holographic model~\cite{Liu:2017frj}, the string model~\cite{Sonnenschein:2017ylo}, the harmonic oscillator based model~\cite{Santopinto:2018ljf}, the light-front quark model~\cite{Chua:2019yqh}, the relativized potential quark model~\cite{Jia:2020vek}, the non-relativistic potential model~\cite{Luo:2023sra}, the relativistic flux tube model~\cite{Jakhad:2023mni} and other quark models~\cite{Karliner:2017kfm,Ortiz-Pacheco:2020hmj}.

On the other hand, the pentaquark interpretation of $\Omega_c$ states has been investigated in the framework of the QCD sum rules~\cite{Wang:2018alb,Wang:2021cku}, the chiral quark model~\cite{Yang:2017rpg,Huang:2017dwn}, the constituent quark model~\cite{An:2017lwg}, the diquark-diquark-antiquark model~\cite{Anisovich:2017aqa}, the heavy-quark spin symmetry model~\cite{Nieves:2017jjx}, the one boson exchange model~\cite{Liu:2018bkx}, the vector meson exchange model~\cite{Montana:2018teh,Montana:2017kjw}, the meson-baryon interaction model~\cite{Ramos:2020bgs}, the chiral quark-soliton model~\cite{Praszalowicz:2022hcp}, the extended local hidden gauge approach~\cite{Debastiani:2017ewu,Debastiani:2018adr}, the Bethe–Salpeter formalism~\cite{Wang:2017smo}, the effective Lagrangian approach~\cite{Huang:2018wgr} and the quasipotential Bethe-Salpeter equation approach~\cite{Zhu:2022fyb}.

Very recently, two new excited states, $\Omega_c^0(3185)$ and $\Omega_c^0(3327)$ are observed in the $\Xi_c^+ K^-$ invariant-mass spectrum by the LHCb Collaboration~\cite{LHCb:2023zpu}.
Bisides, the five narrow $\Omega_c$ states obtained before~\cite{LHCb:2017uwr} are also confirmed.
The measured masses and widths of the two newly found states are
\begin{align}
	M_{\Omega_{c}(3185)} & = 3185.1 \pm 1.7_{-0.9}^{+7.4} \pm 0.2 \mathrm{MeV},  \nonumber \\
	\Gamma_{\Omega_{c}(3185)} & = 50 \pm 7_{-20}^{+10} \mathrm{MeV},  \nonumber \\
	M_{\Omega_{c}(3327)} & = 3327.1 \pm 1.2_{-1.3}^{+0.1} \pm 0.2 \mathrm{MeV},  \nonumber \\
	\Gamma_{\Omega_{c}(3327)} & = 20 \pm 5_{-1}^{+13} \mathrm{MeV}.
\end{align}
So far, there have been a few studies on the two newly discovered states.
In Ref.~\cite{Luo:2023sra}, in the framework of the non-relativistic potential model with Gaussian Expansion Method, the authors' results imply
that the $\Omega_c^0(3327)$ is a good candidate of $\Omega_c^0(1D)$ state with $J^P=5/2^+$.
In Ref.~\cite{Yu:2023bxn}, the $^3P_0$ model calculation results support assigning the observed $\Omega_c^0(3185)$ and $\Omega_c^0(3327)$ as the $2S(3/2^+$) and $1D(3/2^+$) states, respectively.
In Ref.~\cite{Wang:2023wii}, using the QCD sum rules, the $\Omega_c^0(3327)$ is assigned to be $D$-wave $\Omega_c$ state with $J^P=1/2^+, 3/2^+$ or $5/2^+$.
In Ref.~\cite{Feng:2023ixl}, $\Omega_c^0(3185)$ and $\Omega_c^0(3327)$ are studied in the effective Lagrangian approach by assuming they are molecular states.
The results support $\Omega_c^0(3327)$ as a $J^P=3/2^-$ $D^* \Xi$ molecular state, and the $\Omega_c^0(3185)$ may be a meson-baryon molecule with a big $D \Xi$ component.
In Ref.~\cite{Karliner:2023okv}, the assignment of $\Omega_c^0(3185)$ and $\Omega_c^0(3327)$ to $2S_{1/2}$ and $2S_{3/2}$ is discussed.
In Ref.~\cite{Yan:2023ttx}, within a simple contact-range theory in which the couplings are saturated by light-meson exchanges, $\Omega_c^0(3185)$ and $\Omega_c^0(3327)$ match the masses of $J=1/2$ $\Xi D$ and $J=3/2$ $\Xi D^*$, respectively.
In Ref.~\cite{Jakhad:2023mni}, based on the quark-diquark configuration with relativistic flux tube model, $\Omega_c^0(3185)$ and $\Omega_c^0(3327)$ is assigned to be $\left|2S, 3/2^{+}\right\rangle$ and $\left|1D, 3/2^{+}\right\rangle$.
In Ref.~\cite{Xin:2023gkf}, via the QCD sum rules, the numerical results favor assigning $\Omega_c^0(3185)$ as the $D \Xi$ molecular state with the $J^P=1/2^-$, assigning $\Omega_c^0(3327)$ as the $D^* \Xi$ molecular state with the $J^P=3/2^-$.

In addition to the above theoretical methods, quark delocalization color screening model (QDCSM) is a reliable approach, which was developed in the 1990s with the aim of explaining the similarities between nuclear and molecular forces~\cite{Wu:1996fm}.
The model gives a good description of $NN$ and $YN$ interactions and the properties of deuteron~\cite{Ping:2000dx,Ping:1998si,Wu:1998wu,Pang:2001xx}.
It is also employed to calculate the baryon-baryon and baryon-meson scattering phase shifts, and the exotic hadronic states are also studied in this model.
Studies show that color screening is an effective description of the hidden-color channel coupling~\cite{ChenLZ,Huang:2011kf}.
So it is feasible and meaningful to extend this model to investigate the pentaquark interpretation of excited $\Omega_{c}$ states.

In this work, we systematically investigate the $ssc\bar{q}q$ systems in order to find out if there are $\Omega_{c}$ states that are possible to be interpreted as pentaquark states.
The five-body system is calculated by means of the resonating group method to search for bound states.
The strong decay channels of the $ssc\bar{q}q$ systems are investigated to determine the resonance states, based on the conservation of the quantum numbers and the limit of phase space.
In order to ensure the reliability and stability of the calculation results, the parameters used in this work are the same as those used in the previous work~\cite{Yan:2022nxp}.

This paper is organized as follows.
After introduction, the details of QDCSM are presented in section II.
The calculation of the bound state and scattering phase shift is presented in Section III, along with the discussion and analysis of the results.
Finally, the paper ends with summary in Section IV.

\section{QUARK DELOCALIZATION COLOR SCREENING MODEL (QDCSM)}
Herein, the QDCSM is employed to investigate the properties of $ssc\bar{q}q$ systems.
The QDCSM is an extension of the native quark cluster model~\cite{DeRujula:1975qlm,Isgur:1978xj,Isgur:1978wd,Isgur:1979be}.
It has been developed to address multi-quark systems.
The detail of the QDCSM can be found in Refs.~\cite{Wu:1996fm,Huang:2011kf,ChenLZ,Ping:1998si,Wu:1998wu,Pang:2001xx,Ping:2000cb,Ping:2000dx,Ping:2008tp}.
In this sector, we mainly introduce the salient features of this model.
The general form of the pentaquark Hamiltonian is given by
	\begin{align}
		H=&\sum_{i=1}^5\left(m_i+\frac{\boldsymbol{p}_{i}^{2}}{2m_i}\right)-T_{CM} +\sum_{j>i=1}^5 V(\boldsymbol{r}_{ij}),
	\end{align}
where $m_i$ is the quark mass, $\boldsymbol{p}_{i}$ is the momentum of the quark, and $T_{CM}$ is the center-of-mass kinetic energy.
The dynamics of the pentaquark system is driven by a two-body potential
	\begin{align}
		V(\boldsymbol{r}_{ij})= & V_{CON}(\boldsymbol{r}_{ij})+V_{OGE}(\boldsymbol{r}_{ij})+V_{\chi}(\boldsymbol{r}_{ij}).
	\end{align}
The most relevant features of QCD at its low energy regime: color confinement ($V_{CON}$), perturbative one-gluon exchange interaction ($V_{OGE}$), and dynamical chiral symmetry breaking ($V_{\chi}$) have been taken into consideration.

Here, a phenomenological color screening confinement potential ($V_{CON}$) is used as
\begin{align}
	V_{CON}(\boldsymbol{r}_{ij}) = & -a_{c}\boldsymbol{\lambda}_{i}^{c} \cdot \boldsymbol{\lambda}_{j}^{c}\left[  f(\boldsymbol{r}_{ij})+V_{0}\right],
\end{align}
\begin{align}
	f(\boldsymbol{r}_{ij}) =& \left\{\begin{array}{l}
		\boldsymbol{r}_{i j}^{2}, ~~~~~~~~~~~~~ ~i,j ~\text {occur in the same cluster } \\
		\frac{1-e^{-\mu_{q_{i}q_{j}} \boldsymbol{r}_{i j}^{2}}}{\mu_{q_{i}q_{j}}},  ~~~i,j ~\text {occur in different cluster }
	\end{array}\right.   \nonumber
\end{align}
where $a_c$, $V_{0}$ and $\mu_{q_{i}q_{j}}$ are model parameters, and $\boldsymbol{\lambda}^{c}$ stands for the SU(3) color Gell-Mann matrices.
Among them, the color screening parameter $\mu_{q_{i}q_{j}}$ is determined by fitting the deuteron properties, nucleon-nucleon scattering phase shifts, and hyperon-nucleon scattering phase shifts, respectively, with $\mu_{qq}=0.45~$fm$^{-2}$, $\mu_{qs}=0.19~$fm$^{-2}$ and $\mu_{ss}=0.08~$fm$^{-2}$, satisfying the relation, $\mu_{qs}^{2}=\mu_{qq}\mu_{ss}$~\cite{ChenM}.
Besides, we found that the heavier the quark, the smaller this parameter $\mu_{q_{i}q_{j}}$.
When extending to the heavy quark system, the hidden-charm pentaquark system, we took $\mu_{cc}$ as a adjustable parameter from $0.01~$fm$^{-2}$ to $0.001~$fm$^{-2}$, and found that the results were insensitive to the value of $\mu_{cc}$~\cite{HuangPc1}.
Moreover, the $P_{c}$ states were well predicted in the work of Refs.~\cite{HuangPc1,HuangPc2}.
So here we take $\mu_{cc}=0.01~$fm$^{-2}$ and $\mu_{qc}=0.067~$fm$^{-2}$, also satisfy the relation $\mu_{qc}^{2}=\mu_{qq}\mu_{qc}$.

In the present work, we mainly focus on the low-lying negative parity $ssc\bar{q}q$ pentaquark states of $S$-wave, so the spin-orbit and tensor interactions are not included.
The one-gluon exchange potential ($V_{OGE}$), which includes coulomb and chromomagnetic interactions, is written as
	\begin{align}
		V_{OGE}(\boldsymbol{r}_{ij})= &\frac{1}{4}\alpha_{s_{q_i q_j}} \boldsymbol{\lambda}_{i}^{c} \cdot \boldsymbol{\lambda}_{j}^{c}  \\
		&\cdot \left[\frac{1}{r_{i j}}-\frac{\pi}{2} \delta\left(\mathbf{r}_{i j}\right)\left(\frac{1}{m_{i}^{2}}+\frac{1}{m_{j}^{2}}+\frac{4 \boldsymbol{\sigma}_{i} \cdot \boldsymbol{\sigma}_{j}}{3 m_{i} m_{j}}\right)\right],   \nonumber \label{Voge}
	\end{align}
where $\boldsymbol{\sigma}$ is the Pauli matrices and $\alpha_{s_{q_i q_j}}$ is the quark-gluon coupling constant.

However, the quark-gluon coupling constant between quark and anti-quark, which offers a consistent description of mesons from light to heavy-quark sector, is determined by the mass differences between pseudoscalar mesons (spin-parity $J^P=0^-$) and vector (spin-parity $J^P=1^-$), respectively.
For example, from the model Hamiltonian, the mass difference between $D$ and $D^*$ is determined by the chromomagnetic interaction in Eq.~(\ref{Voge}), so the parameter $\alpha_{s_{qc}}$ is determined by fitting the mass difference between $D$ and $D^*$.

The dynamical breaking of chiral symmetry results in the SU(3) Goldstone boson exchange interactions appear between constituent light quarks $u, d$ and $s$.
Hence, the chiral interaction is expressed as
\begin{align}
	V_{\chi}(\boldsymbol{r}_{ij})= & V_{\pi}(\boldsymbol{r}_{ij})+V_{K}(\boldsymbol{r}_{ij})+V_{\eta}(\boldsymbol{r}_{ij}).
\end{align}
Among them
\begin{align}
V_{\pi}\left(\boldsymbol{r}_{i j}\right) =&\frac{g_{c h}^{2}}{4 \pi} \frac{m_{\pi}^{2}}{12 m_{i} m_{j}} \frac{\Lambda_{\pi}^{2}}{\Lambda_{\pi}^{2}-m_{\pi}^{2}} m_{\pi}\left[Y\left(m_{\pi} \boldsymbol{r}_{i j}\right)\right. \nonumber \\
&\left.-\frac{\Lambda_{\pi}^{3}}{m_{\pi}^{3}} Y\left(\Lambda_{\pi} \boldsymbol{r}_{i j}\right)\right]\left(\boldsymbol{\sigma}_{i} \cdot \boldsymbol{\sigma}_{j}\right) \sum_{a=1}^{3}\left(\boldsymbol{\lambda}_{i}^{a} \cdot \boldsymbol{\lambda}_{j}^{a}\right),
\end{align}
\begin{align}
	V_{K}\left(\boldsymbol{r}_{i j}\right) =&\frac{g_{c h}^{2}}{4 \pi} \frac{m_{K}^{2}}{12 m_{i} m_{j}} \frac{\Lambda_{K}^{2}}{\Lambda_{K}^{2}-m_{K}^{2}} m_{K}\left[Y\left(m_{K} \boldsymbol{r}_{i j}\right)\right. \nonumber \\
	&\left.-\frac{\Lambda_{K}^{3}}{m_{K}^{3}} Y\left(\Lambda_{K} \boldsymbol{r}_{i j}\right)\right]\left(\boldsymbol{\sigma}_{i} \cdot \boldsymbol{\sigma}_{j}\right) \sum_{a=4}^{7}\left(\boldsymbol{\lambda}_{i}^{a} \cdot \boldsymbol{\lambda}_{j}^{a}\right),
\end{align}
\begin{align}
	V_{\eta}\left(\boldsymbol{r}_{i j}\right) =&\frac{g_{c h}^{2}}{4 \pi} \frac{m_{\eta}^{2}}{12 m_{i} m_{j}} \frac{\Lambda_{\eta}^{2}}{\Lambda_{\eta}^{2}-m_{\eta}^{2}} m_{\eta}\left[Y\left(m_{\eta} \boldsymbol{r}_{i j}\right)\right. \nonumber \\
	&\left.-\frac{\Lambda_{\eta}^{3}}{m_{\eta}^{3}} Y\left(\Lambda_{\eta} \boldsymbol{r}_{i j}\right)\right]\left(\boldsymbol{\sigma}_{i} \cdot \boldsymbol{\sigma}_{j}\right)\left[\cos \theta_{p}\left(\boldsymbol{\lambda}_{i}^{8} \cdot \boldsymbol{\lambda}_{j}^{8}\right)\right.  \nonumber \\
	&\left.-\sin \theta_{p}\left(\boldsymbol{\lambda}_{i}^{0} \cdot \boldsymbol{\lambda}_{j}^{0}\right)\right],
\end{align}
where $Y(x) = e^{-x}/x$ is the standard Yukawa function.
The physical $\eta$ meson is considered by introducing the angle $\theta_{p}$ instead of the octet one. The $\boldsymbol{\lambda}^a$ are the SU(3) flavor Gell-Mann matrices.
The values of $m_\pi$, $m_k$ and $m_\eta$ are the masses of the SU(3) Goldstone bosons, which adopt the experimental values~\cite{ParticleDataGroup:2020ssz}.
The chiral coupling constant $g_{ch}$, is determined from the $\pi N N$ coupling constant through
\begin{align}
	\frac{g_{c h}^{2}}{4 \pi} & = \left(\frac{3}{5}\right)^{2} \frac{g_{\pi N N}^{2}}{4 \pi} \frac{m_{u, d}^{2}}{m_{N}^{2}}.
\end{align}
Assuming that flavor SU(3) is an exact symmetry, it will only be broken by the different mass of the strange quark.
The other symbols in the above expressions have their usual meanings.
All the parameters shown in Table~\ref{parameters} are fixed by masses of the ground baryons and mesons. Table~\ref{hadrons} shows the masses of the baryons and mesons used in this work.

In the QDCSM, quark delocalization was introduced to enlarge the model variational space to take into account the mutual distortion or the internal excitations of nucleons in the course of interaction.
It is realized by specifying the single particle orbital wave function of the QDCSM as a linear combination of left and right Gaussians, the single particle orbital wave functions used in the ordinary quark cluster model
\begin{eqnarray}
	\psi_{\alpha}(\boldsymbol {S_{i}} ,\epsilon) & = & \left(
	\phi_{\alpha}(\boldsymbol {S_{i}})
	+ \epsilon \phi_{\alpha}(-\boldsymbol {S_{i}})\right) /N(\epsilon), \nonumber \\
	\psi_{\beta}(-\boldsymbol {S_{i}} ,\epsilon) & = &
	\left(\phi_{\beta}(-\boldsymbol {S_{i}})
	+ \epsilon \phi_{\beta}(\boldsymbol {S_{i}})\right) /N(\epsilon), \nonumber \\
	N(S_{i},\epsilon) & = & \sqrt{1+\epsilon^2+2\epsilon e^{-S_i^2/4b^2}}. \label{1q}
\end{eqnarray}
It is worth noting that the mixing parameter $\epsilon$ is not an adjusted one but determined variationally by the dynamics of the multi-quark system itself.
In this way, the multi-quark system chooses its favorable configuration in the interacting process.
This mechanism has been used to explain the cross-over transition between hadron phase and quark-gluon plasma phase~\cite{Xu}.

In addition, the dynamical calculation is carried out using the resonating group method and the generating coordinates method.
The details of the two methods can be seen in Appendix A, and the way of constructing wave functions are presented in Appendix B.

\begin{table}[ht]
	\caption{\label{parameters}Model parameters used in this work:
		$m_{\pi}=0.7$ fm$^{-1}$,
		$m_{K}=2.51$ fm$^{-1}$,
		$m_{\eta}=2.77$ fm$^{-1}$,
		$\Lambda_{\pi}=4.2$ fm$^{-1}$,
		$\Lambda_{K}=5.2$ fm$^{-1}$,
		$\Lambda_{\eta}=5.2$ fm$^{-1}$,
		$g_{ch}^2/(4\pi)$=0.54.}
	\begin{tabular}{cccccc} \hline\hline
		~~~~$b$~~~~ & ~~$m_{q}$~~ & ~~~$m_{c}$~~~  & ~~~$V_{0_{qq}}$~~~~&~~~$V_{0_{q\bar{q}}}$~~~~& ~~~$ a_c$~~~   \\
		(fm) & (MeV) & (MeV)  & (fm$^{-2}$)  & (fm$^{-2}$) & ~(MeV\,fm$^{-2}$)~  \\
		0.518  & 313 & 1788   &    -1.288 &  -0.743  &  58.03 \\ \hline
		$\alpha_{s_{qs}}$ &  $\alpha_{s_{qc}}$ &  $\alpha_{s_{sc}}$ & $\alpha_{s_{q\bar{q}}}$ & $\alpha_{s_{s\bar{q}}}$ & $\alpha_{s_{c\bar{q}}}$  \\
		0.524 & 0.467  & 0.351 & 1.491 & 1.423 & 1.200   \\ \hline\hline
	\end{tabular}
\end{table}

\begin{table}[ht]
	\caption{The masses (in MeV) of the baryons and mesons. Experimental values are taken from the Particle Data Group (PDG)~\cite{ParticleDataGroup:2020ssz}.}
	\begin{tabular}{cccc}
		\hline \hline
		~~~~Hadron~~~~ & ~~~~~$I(J^P)$~~~~~  & ~~~~~~$M_{Exp}$~~~~~~ & ~~~~$M_{Theo}$~~~~  \\ \hline
		$N$              & $1/2(1/2^+)$ & 939   & 939  \\
		$\Delta$         & $3/2(3/2^+)$ & 1232  & 1232 \\
		$\Sigma_c$       & $1(1/2^+)$   & 2455  & 2465 \\
		$\Sigma^*_c$     & $1(3/2^+)$   & 2490  & 2518 \\
		$\Lambda_c$      & $0(1/2^+)$   & 2286  & 2286 \\
		$\Xi      $      & $1/2(1/2^+)$ & 1318  & 1375 \\
		$\Xi^*    $      & $1/2(3/2^+)$ & 1536  & 1496 \\
		$\Xi_c    $      & $1/2(1/2^+)$ & 2467  & 2551 \\
		$\Xi_c^{\prime}$ & $1/2(1/2^+)$ & 2577  & 2621 \\
		$\Xi_c ^* $      & $1/2(3/2^+)$ & 2645  & 2638 \\
     	$\Omega_c $      & $0(1/2^+)$   & 2695  & 2785 \\
		$\Omega_c^* $    & $0(3/2^+)$   & 2766  & 2796 \\
		$\pi$            & $1(0^-)$     & 139   & 139  \\
		$\rho$           & $1(1^-)$     & 770   & 770  \\
		$\omega$         & $0(1^-)$     & 782   & 722  \\
		$\bar{K}$        & $1/2(0^-)$   & 495   & 495  \\
		$\bar{K}^*$      & $1/2(1^-)$   & 892   & 814  \\
		$D$              & $1/2(0^-)$   & 1869  & 1869 \\
		$D^*$            & $1/2(1^-)$   & 2007  & 1952  \\  \hline\hline
	\end{tabular}
	\label{hadrons}
\end{table}

\section{The results and discussions}
In this work, we investigate the $S-$wave $ssc\bar{q}q$ pentaquark systems in the framework of QDCSM with resonating group method.
The quantum numbers of the pentaquark system are $I=0$, $J^P = 1/2^-, 3/2^-$ and $5/2^-$.
Three structures $qss-\bar{q}c$, $qsc-\bar{q}s$ and $ssc-\bar{q}q$, as well as the coupling of these structures are taken into consideration.
To find out if there exists any bound state, we carry out a dynamic bound-state calculation.
The scattering process is also studied to obtain the genuine resonance state.
The introduction of the bound state calculation and scattering process can be seen in Appendix A.
Moreover, the calculation of root mean square (RMS) of cluster spacing is helpful to explore the structure of the bound state or resonance state on the one hand, and to further estimate whether the observed states are resonance state or scattering state on the other hand.

The single-channel results of different systems are listed in Tables~\ref{sc 1/2}, \ref{sc 3/2} and \ref{sc 5/2}, respectively.
The first column headed with Structure incluedes $qss-\bar{q}c$, $qsc-\bar{q}s$ and $ssc-\bar{q}q$ three kinds.
The second and third columns, headed with $\chi^{f_i}$ and $\chi^{\sigma_{j}}$, denote the way how wave functions constructed, which can be seen in Appendix B.
The forth column headed with Channel gives the physical channels involved in the present work.
The fifth column headed with $E_{th}^{Theo}$ refers to the theoretical value of non-interacting baryon-meson threshold.
The sixth column headed with $E_{sc}$ shows the energy of each single channel.
The values of binding energies $E_B$= $E_{sc} -E_{th}^{Theo}$ are listed in the eighth column only if $E_B<0$ MeV.
Finally, the experimental thresholds $E_{th}^{Exp}$ (the sum of the experimental masses of the corresponding baryon and meson) along with corrected energies $E^{\prime} = E_{th}^{Exp} + E_B$ are given in last two columns.

As for coupled-channel, the results are listed in the Tables~\ref{cc 1/2}, \ref{cc 3/2} and \ref{cc 5/2}, respectively.
The first column represents the structures involved in the channel coupling and the second column is the theoretical value of the lowest threshold.
The third column headed with $E_{cc}$ shows the energy of coupled-channel.
The definitions of $E_B$, $E_{th}^{Exp}$ and $E^\prime$ in coupled-channel calculation are similar to their definitions in single-channel calculation.

\subsection{$J^P=\frac{1}{2}^-$ sector}

First of all, an intuitive analysis can be based on the results of single-channel calculation, which is shown in the Table~\ref{sc 1/2}.
Except for the $\Omega_c \omega$ and $\Omega_c^* \omega$, the energies of other single-channels are all higher than the corresponding thresholds.
The binding energies of the $\Omega_c \omega$ and $\Omega_c^* \omega$ are -3 MeV and -4 MeV, respectively.
Since that different channels of the system are influenced by each other, so it is unavoidable to take into account the channel coupling effect.

\begin{table*}[htb]
	\caption{\label{sc 1/2}The single-channel energies of the $ssc\bar{q}q$ pentaquark system with $J^P=\frac{1}{2}^-$ (unit: MeV).}
	\begin{tabular}{c c c c c c c c c} \hline\hline
		
		~~~Structure~~~&~~~~$\chi^{f_i}$~~~~ & ~~~~~~$\chi^{\sigma_j}$~~~~~~ & ~Channel~ & ~~~~~~$E_{th}^{Theo}$~~~~~~ & ~~~~$E_{sc}$~~~~ & ~~~~~$E_{B}$~~~~~ &  ~~~~~$E_{th}^{Exp}$~~~~~ & ~~~~~$E'$~~~~~   \\ \hline
		
		$qss-\bar{q}c$ & $i=2$ & $j=1$ & $\Xi D$              & 3235 & 3238 & ub & 3187 & 3190  \\
		               & $i=2$ & $j=2$ & $\Xi D^*$            & 3319 & 3321 & ub & 3325 & 3327  \\
		               & $i=2$ & $j=3$ & $\Xi^* D^*$          & 3441 & 3447 & ub & 3543 & 3549  \\
		
		$qsc-\bar{q}s$ & $i=2$ & $j=1$ & $\Xi^{\prime}_c \bar{K}$   & 3130 & 3137 & ub & 3072 & 3079  \\
		               & $i=2$ & $j=1$ & $\Xi_c \bar{K}$            & 3060 & 3066 & ub & 2962 & 2968  \\
		               & $i=2$ & $j=2$ & $\Xi^{\prime}_c \bar{K}^*$ & 3449 & 3454 & ub & 3469 & 3574  \\
		               & $i=2$ & $j=2$ & $\Xi_c \bar{K}^*$          & 3379 & 3386 & ub & 3359 & 3366  \\
		               & $i=2$ & $j=3$ & $\Xi^*_c \bar{K}^*$        & 3466 & 3472 & ub & 3537 & 3543  \\
		
		$ssc-\bar{q}q$ & $i=1$ & $j=2$ & $\Omega_c \omega$    & 3548 & 3545 & -3 & 3477 & 3474  \\
		               & $i=1$ & $j=3$ & $\Omega_c^* \omega$  & 3558 & 3554 & -4 & 3548 & 3544  \\ \hline\hline
	\end{tabular}
\end{table*}

In order to better understand channel coupling, we first couple the channels with the same spatial structure, and then coupled the channels with different spatial structures.
By solving the Schrodinger equation with channel coupling, we can obtain a series of eigenvalues theoretically.
Only the lowest energy is presented in Table~\ref{cc 1/2}, because whether the system can form a bound state depends on whether the lowest energy is below the lowest threshold.
After coupling the channels with the same spatial structure, the loweset energiges of $qss-\bar{q}c$ and $qsc-\bar{q}s$ systems are still higher than their respective thresholds.
Besides, the $ssc-\bar{q}q$ system forms a bound state with the binding energy of 3 MeV.

Further, we couple the channels with two different spatial structures, $qss-\bar{q}c$ and $ssc-\bar{q}q$, and then add the third spatial structure $ssc-\bar{q}q$ into the coupling.
The result shows that the coupling of $qss-\bar{q}c$ and $ssc-\bar{q}q$ depresses the lowest energy and makes it 5 MeV below the threshold $\Xi D$.
After an overall coupling of all channels, the lowest energy of the system is still higher than the lowest threshold of channel $\Xi_c \bar{K}$, indicating that the $J^P=1/2^-$ $ssc\bar{q}q$ pentaquark system does not form a genuine bound state.

\begin{table*}[htb]
	\caption{\label{cc 1/2}The coupled-channel energies of the $ssc\bar{q}q$ pentaquark system with $J^P=\frac{1}{2}^-$ (unit: MeV).}
	\begin{tabular}{c c c c c c} \hline\hline
		
		~~~~Coupled-structure~~~~& ~~~~~~~$E_{th}^{Theo}$ (Channel)~~~~~~~ & ~~~~~~$E_{cc}$~~~~~~ & ~~~~~~~$E_{B}$~~~~~~~ &  ~~~~~~$E_{th}^{Exp}$~~~~~~ & ~~~~~$E'$~~~~~   \\ \hline
		
		$qss-\bar{q}c$                             & 3235 ($\Xi D$)           & 3237 & ub & 3187 & 3190  \\
		$qsc-\bar{q}s$                             & 3060 ($\Xi_c \bar{K}$)   & 3065 & ub & 2962 & 2967  \\
		$ssc-\bar{q}q$                             & 3548 ($\Omega_c \omega$) & 3545 & -3 & 3477 & 3474  \\
		$qss-\bar{q}c,~ssc-\bar{q}q$               & 3235 ($\Xi D$)           & 3230 & -5 & 3187 & 3192  \\
		$qss-\bar{q}c,~qsc-\bar{q}s,~ssc-\bar{q}q$ & 3060 ($\Xi_c \bar{K}$)   & 3064 & ub & 2962 & 2966  \\ \hline\hline
	\end{tabular}
\end{table*}

According to the results above, some quasi-bound states are obtained in the single-channel calculation and structure coupling.
By coupling to open channels, these states can decay to the corresponding open channels and may become resonance states.
Yet it is not excluded that these states become scattered states under the coupling effect of open channels and closed channels.
So to determine whether resonance states would exist, we continue to study the scattering phase shifts of possible open channels in the QDCSM.
The resonance masses and the decay widths of the resonance states are also calculated.
The current calculation applies only to the decay of $S-$wave open channels.

\begin{figure}[htb]
	\centering
	\includegraphics[width=8cm]{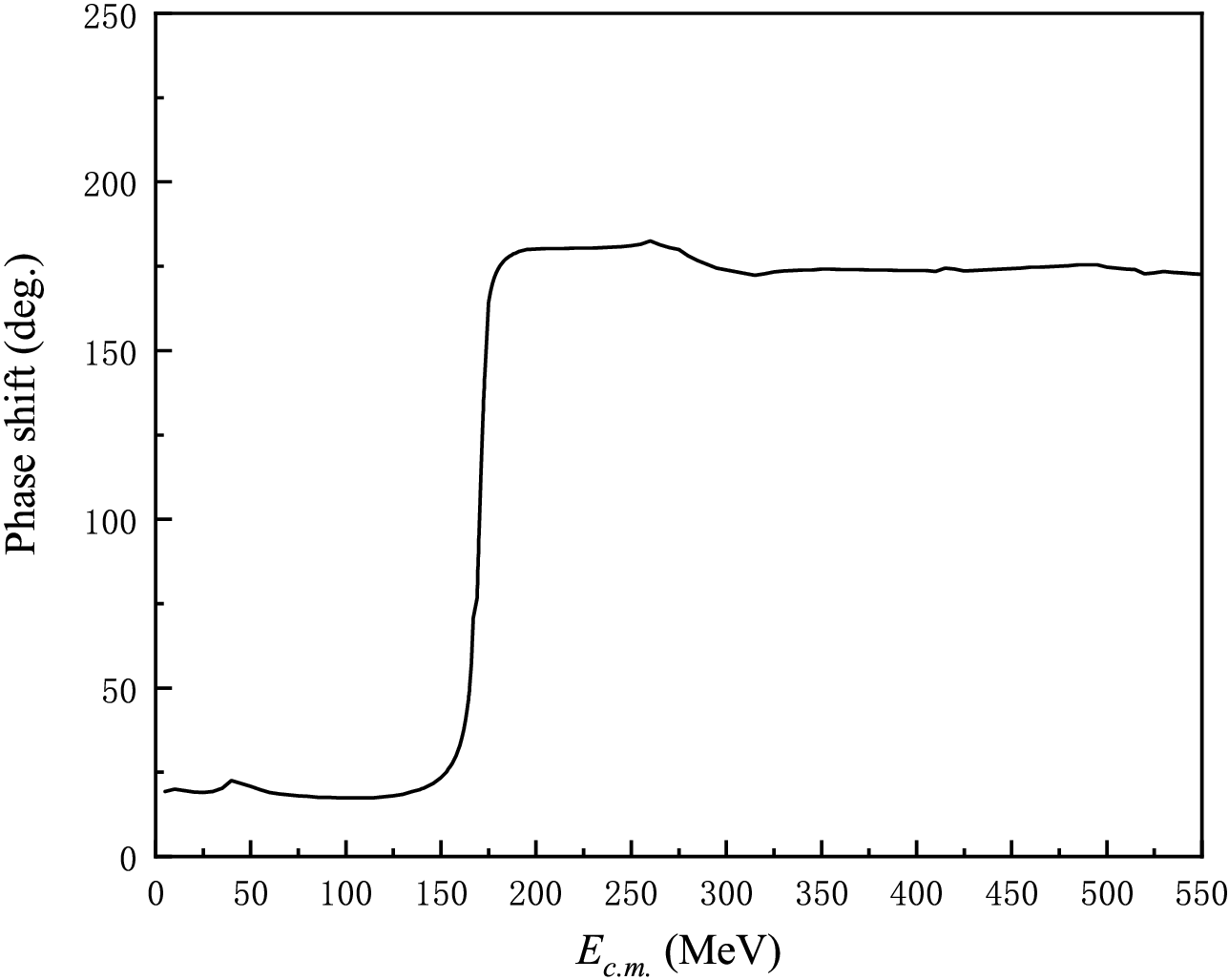}\
	\caption{\label{shift1}  The phase shift of the open channel $\Xi_c \bar{K}$ with $J^P=\frac{1}{2}^-$.}
\end{figure}

\begin{figure}[htb]
	\centering
	\includegraphics[width=8cm]{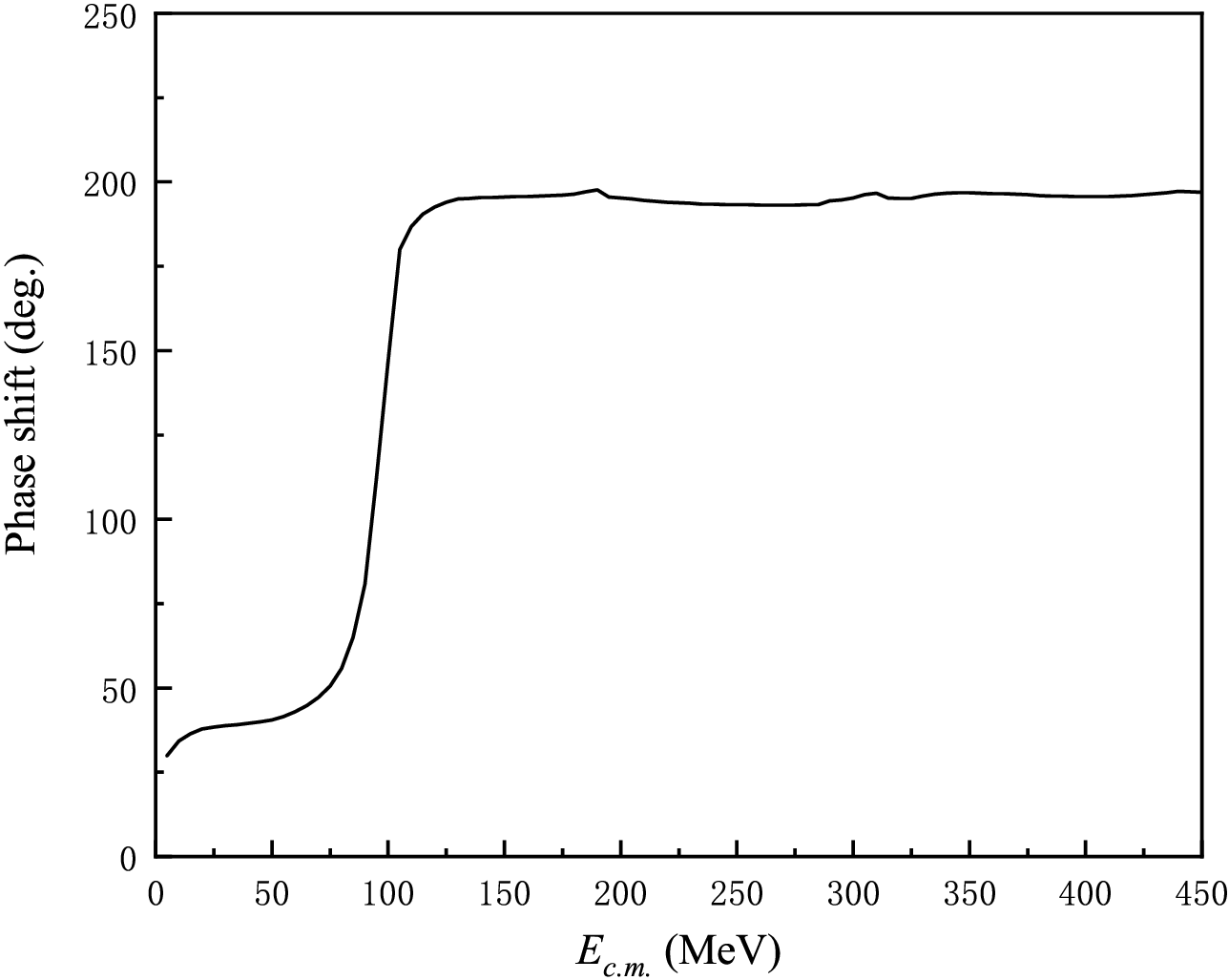}\
	\caption{\label{shift2}  The phase shift of the open channel $\Xi_c^{\prime} \bar{K}$ with $J^P=\frac{1}{2}^-$.}
\end{figure}

First, in order to determine whether $\Xi D$ forms a resonance state, we study the scattering process of open channels $\Xi_c \bar{K}$ and $\Xi_c^\prime \bar{K}$, because the two channels have lower thresholds than the energy of the $\Xi D$ state.
The phase shifts of $\Xi_c \bar{K}$ and $\Xi_c^\prime \bar{K}$ are shown in Fig.~\ref{shift1} and Fig.~\ref{shift2}, respectively.
It is obvious that both phase shifts show a sharp increase around the corresponding resonance mass, which indicates that the $\Xi D$ state becomes a resonance state in both $\Xi_c \bar{K}$ and $\Xi_c^\prime \bar{K}$ scattering process.
The resonance mass, corrected mass and the decay width are summarized as follows:
\begin{align}
	\text{In}~\Xi_c \bar{K}~ \text{channel}:~~~~ M_{res}^{Theo} &= 3230~ \mathrm{MeV},     \nonumber \\
	                 M_{res}^{\prime} &= 3182~ \mathrm{MeV},     \nonumber \\
	                     \Gamma_{res} &= 8.4~ \mathrm{MeV},      \nonumber \\
	\text{In}~ \Xi_c^\prime \bar{K}~ \text{channel}:~~~~ M_{res}^{Theo} &= 3221~ \mathrm{MeV},     \nonumber \\
		                     M_{res}^{\prime} &= 3174~ \mathrm{MeV},     \nonumber \\
		                         \Gamma_{res} &= 33.6~ \mathrm{MeV}.      \nonumber
\end{align}

Thus, a resonance state dominated by $\Xi D$ with $J^P=1/2^-$ in the decay channel $\Xi_c \bar{K}$ and $\Xi_c^\prime \bar{K}$, with the corrected resonance mass 3174$\sim$3182 MeV and decay width 42 MeV, is confirmed.
This is consistent with the newly reported $\Omega_c(3185)$, the mass and decay width of which are $3185.1 \pm 1.7_{-0.9}^{+7.4} \pm 0.2$ MeV and $50 \pm 7_{-20}^{+10}$ MeV, respectively.
Therefore, in our quark model calculation, the $\Omega_c(3185)$ can be well interpreted as a $\Xi D$ resonance state with $J^P=1/2^-$.
In addition, one may be curious about the cusps around 50 MeV and 250 MeV in the Fig.~\ref{shift1}.
They are caused by the thresholds of the corresponding single channels and the similar situation can also be found in Fig.~\ref{shift2}.

In order to investigate the structure of this $\Xi D$ resonance, we further calculate its RMS.
It is worth noting that, the scattering state has no real RMS since the relative motion wave functions of the scattered states are non-integrable in the infinite space.
If we calculate the RMS of a scattering state in a limited space, we can only obtain a value that increases with the expansion of computing space.
Although the wave function of a resonance state is also non-integrable, we can calculate the RMS of the main component of the resonance state, whose wave function is integrable.
In this way, we can calculate the RMS of various states to identify the nature of these states by keep expanding the computing space.
According to the numerical result, the RMS of the resonance state $\Xi D$ is 1.9 fm, indicating that it is likely to be a molecular state.

$\Omega_c \omega$, $\Omega_c^* \omega$ and their coupling also form quasi-bound states in the previous calculations.
The energies of the $\Omega_c \omega$ and $\Omega_c^* \omega$ single channels are about 485$\sim$495 MeV above the threshold of $\Xi_c \bar{K}$ and 415$\sim$425 MeV above the threshold of $\Xi_c^\prime \bar{K}$.
However, in Fig.~\ref{shift1} and Fig.~\ref{shift2}, the sharp increase structure of phase shift representing the resonance state, does not appear around the energies of $\Omega_c \omega$ and $\Omega_c^* \omega$ channels.
Besides, it is still possible that $\Omega_c \omega$ and $\Omega_c^* \omega$ decay to other open channels than $\Xi_c \bar{K}$ or $\Xi_c^\prime \bar{K}$ channels.

\begin{figure}[htb]
	\centering
	\includegraphics[width=8cm]{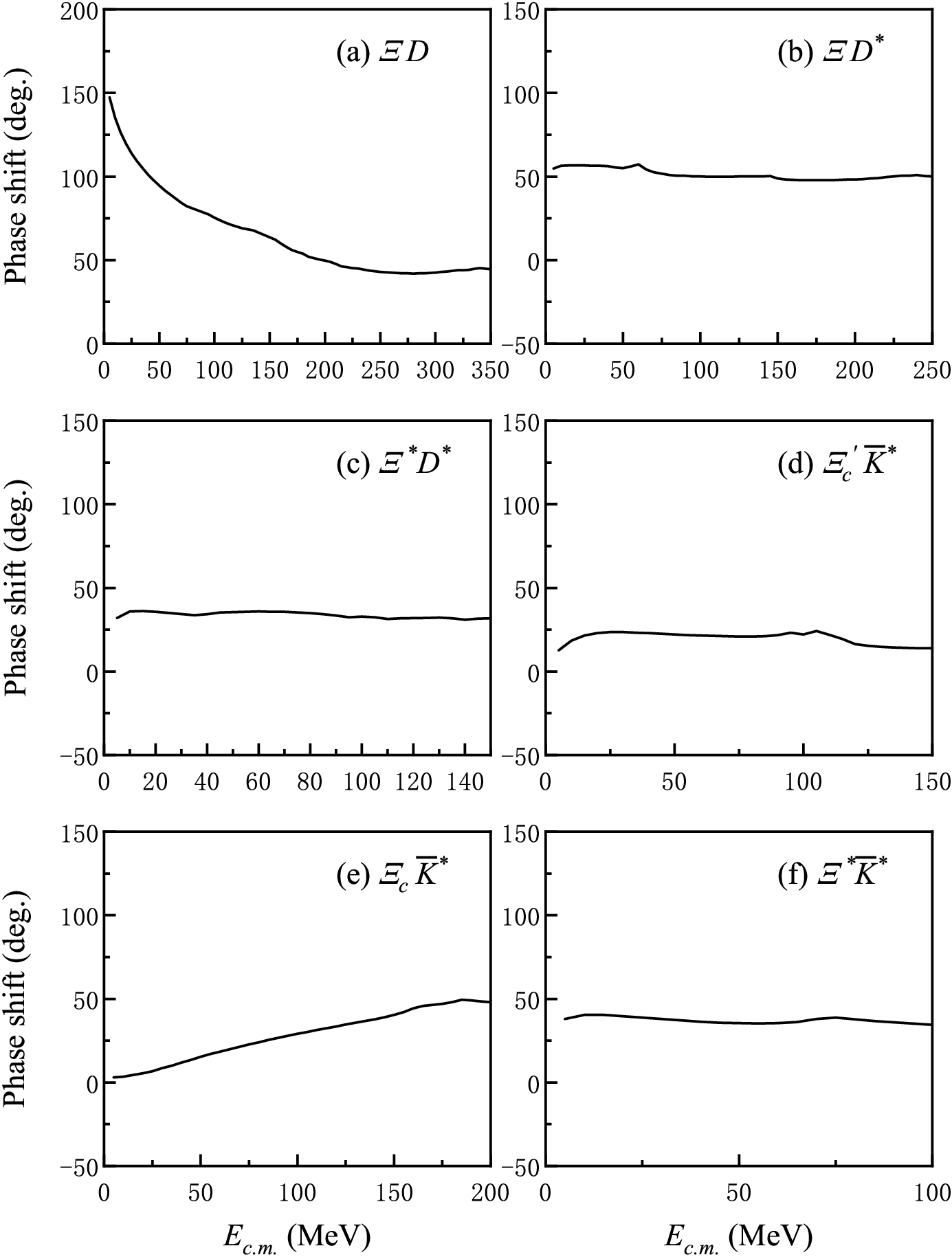}\
	\caption{\label{shift3}  The phase shifts of different open channels with $J^P=\frac{1}{2}^-$.}
\end{figure}

Therefore, we also calculate the phase shifts of other different open channels with channel coupling, which are shown in Fig.~\ref{shift3}.
The ranges of incident energy for different open channels are determined to fit the energy of $\Omega_c^* \omega$, which is the highest energy of the system.
As a result, the ranges of incident energy for different open channels are not the same in Fig.~\ref{shift3}.
After considering the different decay channels, no resonance state of $\Omega_c \omega$ or $\Omega_c^* \omega$ is found.
This can be explained by the effect of channel coupling, which should be fully considered.
As listed in the Table~\ref{cc 1/2}, the energy of $qss-\bar{q}c$ structure coupling is above the corresponding threshold $\Xi D$.
Nevertheless, the energy of $qss-\bar{q}c$ structure is pushed below the threshold $\Xi D$, after being coupled to the $ssc-\bar{q}q$ structure.
As a result, the energy of the previous quasi-bound state $\Omega_c \omega$ is pushed above the corresponding threshold by a reaction, which causes the narrow resonance state to disappear after coupling with the $qss-\bar{q}c$ structure.
The same thing happens in the phase shift of the open channel $\Xi_c^\prime \bar{K}$.
Therefore, the narrow resonance state we just discussed is not a genuine resonance state.

\subsection{$J^P=\frac{3}{2}^-$ sector}

The single-channel energies of $ssc\bar{q}q$ system with $J^P=3/2^-$ are listed in the Table~\ref{sc 3/2}.
Two bound states are obtained in the $\Omega_c \omega$ and $\Omega^{*}_c \omega$ channel, while the energies of other channels are all above the corresponding thresholds.
The binding energies of the $\Omega_c \omega$ and $\Omega^{*}_c \omega$ state are -2 MeV and -4 MeV, respectively.

\begin{table*}[htb]
	\caption{\label{sc 3/2}The single-channel energies of the $ssc\bar{q}q$ pentaquark system with $J^P=\frac{3}{2}^-$ (unit: MeV).}
	\begin{tabular}{c c c c c c c c c} \hline\hline
		
		~~~Structure~~~&~~~~$\chi^{f_i}$~~~~ & ~~~~~~$\chi^{\sigma_j}$~~~~~~ & ~Channel~ & ~~~~~~$E_{th}^{Theo}$~~~~~~ & ~~~~$E_{sc}$~~~~ & ~~~~~$E_{B}$~~~~~ &  ~~~~~$E_{th}^{Exp}$~~~~~ & ~~~~~$E'$~~~~~   \\ \hline
		
		$qss-\bar{q}c$ & $i=2$ & $j=4$ & $\Xi D^*$            & 3319 & 3323 & ub & 3325 & 3329  \\
		& $i=2$ & $j=5$ & $\Xi^* D$            & 3357 & 3362 & ub & 3405 & 3410  \\
		& $i=2$ & $j=6$ & $\Xi^* D^*$          & 3441 & 3446 & ub & 3543 & 3548  \\
		
		$qsc-\bar{q}s$ & $i=2$ & $j=4$ & $\Xi^{\prime}_c \bar{K}^*$ & 3449 & 3457 & ub & 3469 & 3477  \\
		& $i=2$ & $j=4$ & $\Xi_c \bar{K}^*$          & 3379 & 3386 & ub & 3359 & 3366  \\
		& $i=2$ & $j=5$ & $\Xi_c^* \bar{K}$          & 3147 & 3153 & ub & 3140 & 3146  \\
		& $i=2$ & $j=6$ & $\Xi^*_c \bar{K}^*$        & 3466 & 3472 & ub & 3537 & 3543  \\
		
		$ssc-\bar{q}q$ & $i=1$ & $j=4$ & $\Omega_c \omega$    & 3548 & 3546 & -2 & 3477 & 3475  \\
		& $i=1$ & $j=6$ & $\Omega_c^* \omega$  & 3558 & 3554 & -4 & 3548 & 3544  \\ \hline\hline
	\end{tabular}
\end{table*}

For the $ssc\bar{q}q$ system with $J^P=3/2^-$, channel coupling of various structures is also considered, which is listed in Table~\ref{cc 3/2}.
Similar to the previous section with $J^P=1/2^-$, we first carry out the channel coupling with the same spatial structure.
Single structure coupling $qss-\bar{q}c$ and $qsc-\bar{q}s$ are all unbound according to the numerical results.
In addition, the $ssc-\bar{q}q$ structure coupling slightly depresses the energy of $\Omega_{c} \omega$, although it is not numerically significant.

\begin{table*}[htb]
	\caption{\label{cc 3/2}The coupled-channel energies of the $ssc\bar{q}q$ pentaquark system with $J^P=\frac{3}{2}^-$ (unit: MeV).}
	\begin{tabular}{c c c c c c} \hline\hline
		
		~~~~Coupled-structure~~~~& ~~~~~~~$E_{th}^{Theo}$ (Channel)~~~~~~~ & ~~~~~~$E_{cc}$~~~~~~ & ~~~~~~~$E_{B}$~~~~~~~ &  ~~~~~~$E_{th}^{Exp}$~~~~~~ & ~~~~~$E'$~~~~~   \\ \hline
		
		$qss-\bar{q}c$                             & 3319 ($\Xi D^*$)         & 3322 & ub & 3325 & 3328  \\
		$qsc-\bar{q}s$                             & 3147 ($\Xi_c^* \bar{K}$) & 3150 & ub & 3140 & 3143  \\
		$ssc-\bar{q}q$                             & 3548 ($\Omega_c \omega$) & 3546 & -2 & 3477 & 3475  \\
		$qss-\bar{q}c,~ssc-\bar{q}q$               & 3319 ($\Xi D^*$)         & 3321 & ub & 3325 & 3327  \\
		$qss-\bar{q}c,~qsc-\bar{q}s,~ssc-\bar{q}q$ & 3147 ($\Xi_c^* \bar{K}$) & 3145 & -2 & 3140 & 3138  \\ \hline\hline
	\end{tabular}
\end{table*}

\begin{figure}[htb]
	\centering
	\includegraphics[width=8cm]{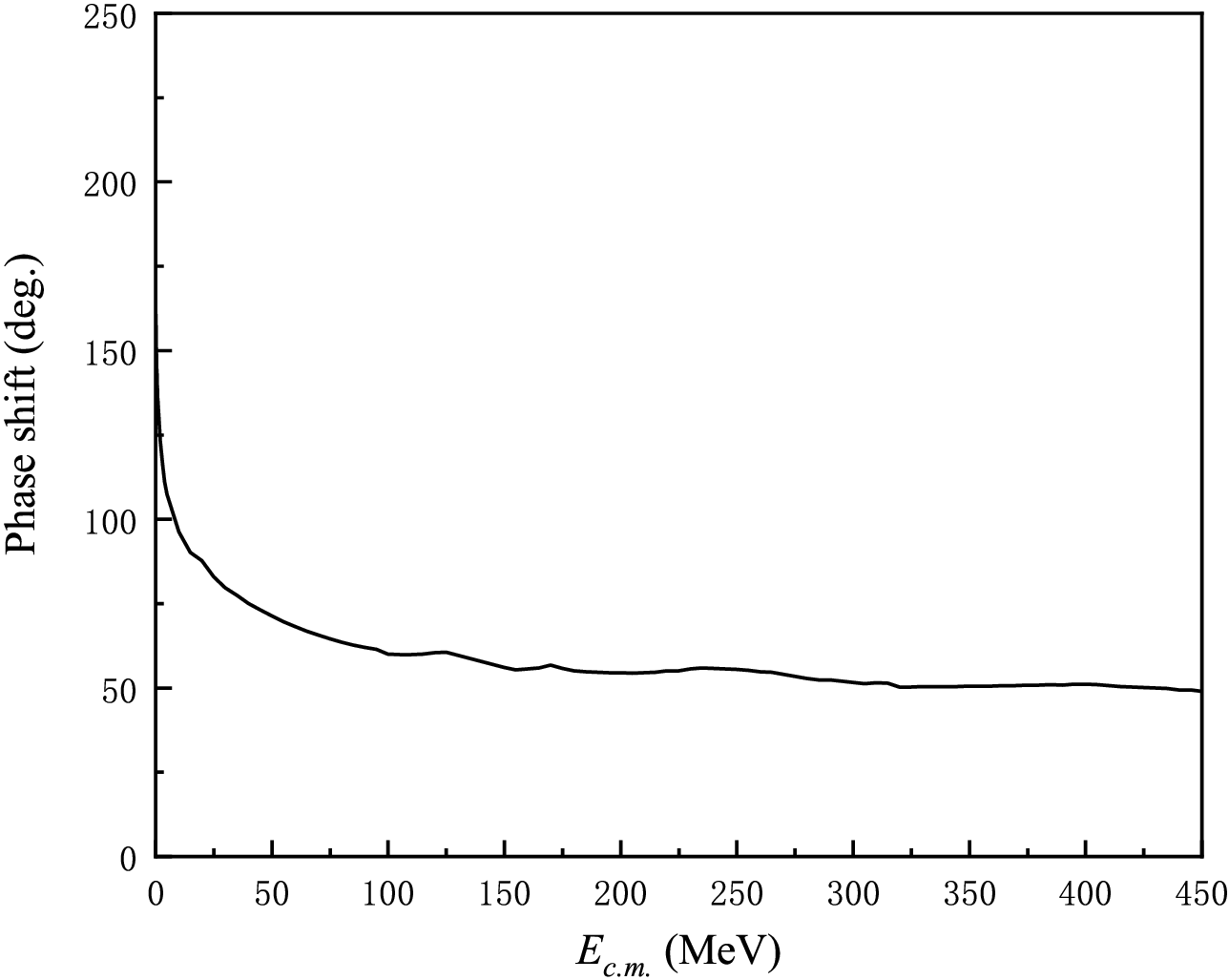}\
	\caption{\label{shift4}  The phase shift of the open channel $\Xi_c^* \bar{K}$ with $J^P=\frac{3}{2}^-$.}
\end{figure}

After calculating two different structure coupling, we continue to add the third structure into the coupling.
As one can see, after coupling all three structures, the lowest energy of the whole system is 2 MeV lower than the energy of the threshold $\Xi_c^* \bar{K}$.
Since the $\Xi_c^* \bar{K}$ is the lowest threshold of the $ssc\bar{q}q$ system with $J^P=3/2^-$, a stable bound state is obtained and its corrected mass is 3138 MeV.
Besides, the bound state conclusion can also be confirmed in the scattering process.
In Fig~\ref{shift4}, as the incident energy approaches 0 MeV, the phase shift of the open channel $\Xi_c^* \bar{K}$ tends to 180 degrees, which conforms to the characteristics of a bound state.
According to the further calculation, this state is dominated by $\Xi_c^* \bar{K}$ and the RMS calculation is 1.8 fm.
The mass is close to the mass of $\Omega_c(3120)$, which is $3119.1 \pm 0.3 \pm 0.9 \pm 0.3$ MeV.
In addition, the bound state $\Xi_c^* \bar{K}$ can still decay to some $D$-wave channels, such as $\Xi_c \bar{K}$, through the tensor force coupling, which is the next step of our research in the future.
However, the decay width of this type of decay is usually very narrow, according to our previous research~\cite{Chen:2011zzb}.
This corresponds to the decay width of $\Omega_c(3120)$, which is $0.60 \pm 0.63$ MeV.
In this case, $\Omega_c(3120)$ could be interpreted as a $\Xi_c^* \bar{K}$ molecular state with $J^P=3/2^-$ in present calculation.

\begin{figure}[htb]
	\centering
	\includegraphics[width=8cm]{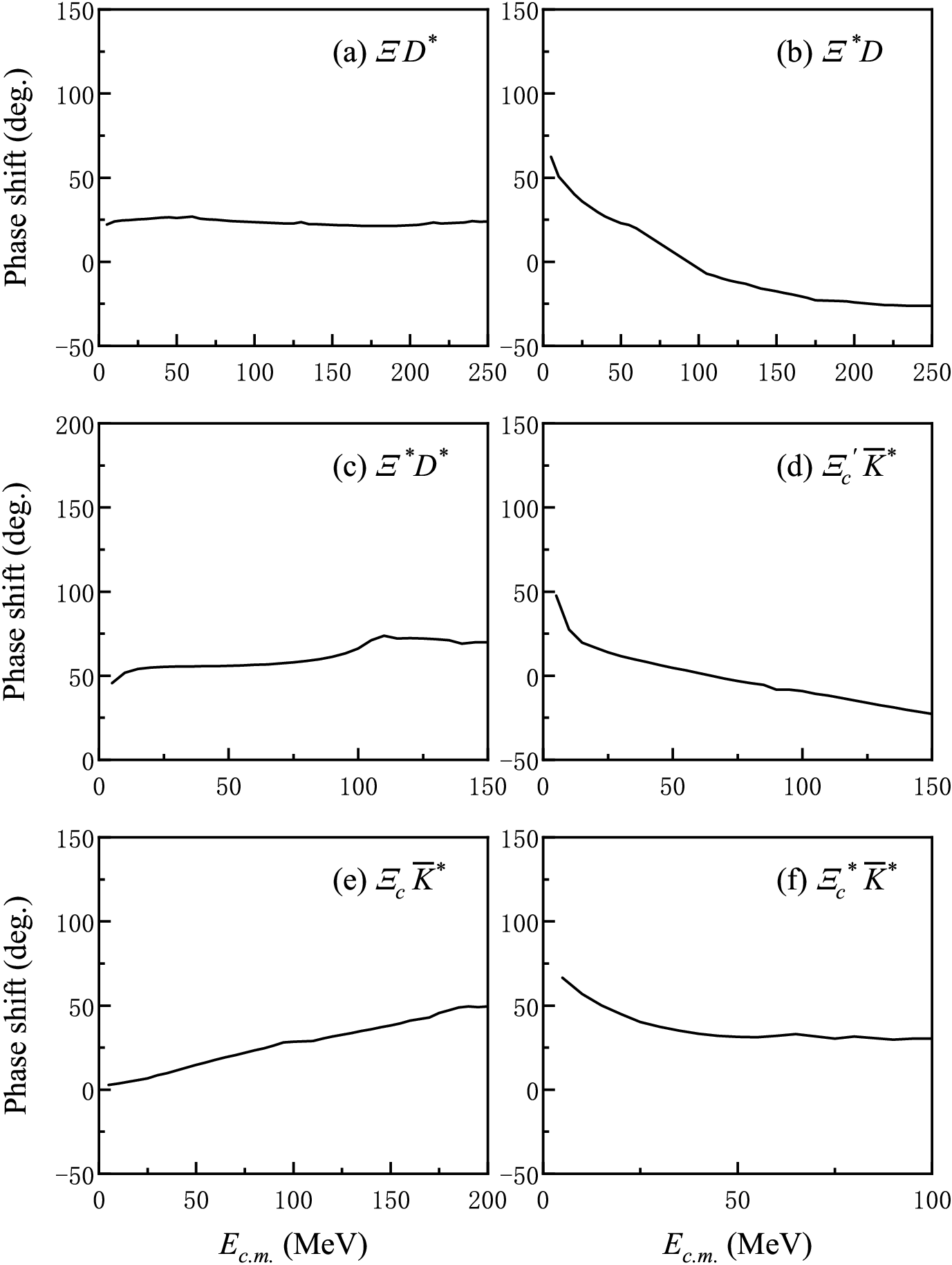}\
	\caption{\label{shift5}  The phase shifts of different open channels with $J^P=\frac{3}{2}^-$.}
\end{figure}

Furthermore, the scattering process is studied to examine whether $\Omega_c \omega$ and $\Omega_c^* \omega$ can form resonance states.
The phase shifts of different $S-$wave open channels with channel coupling are shown in Fig~\ref{shift4} and Fig~\ref{shift5}.
However, the phase shifts of all open channels do not show a sharp increase around the energies of the quasi-bound state $\Omega_c \omega$ or $\Omega_c^* \omega$.
The result shows that the $\Omega_c \omega$ and $\Omega_c^* \omega$ become scattering states rather than resonance states after being coupled to other channels.

\subsection{$J^P=\frac{5}{2}^-$ sector}

For the $ssc\bar{q}q$ system with $J^P=5/2^-$, there is three channels $\Xi^* D^*$, $\Xi_c^* \bar{K}^*$ and $\Omega_{c}^* \omega$.
The energies obtained in the single-channel calculation are presented in Table~\ref{sc 5/2}.
The $\Omega_{c}^* \omega$ forms a bound state, which will be examined later to see if it is a resonance state.

\begin{table*}[htb]
	\caption{\label{sc 5/2}The single-channel energies of the $ssc\bar{q}q$ pentaquark system with $J^P=\frac{5}{2}^-$ (unit: MeV).}
	\begin{tabular}{c c c c c c c c c} \hline\hline
		
		~~~Structure~~~&~~~~$\chi^{f_i}$~~~~ & ~~~~~~$\chi^{\sigma_j}$~~~~~~ & ~Channel~ & ~~~~~~$E_{th}^{Theo}$~~~~~~ & ~~~~$E_{sc}$~~~~ & ~~~~~$E_{B}$~~~~~ &  ~~~~~$E_{th}^{Exp}$~~~~~ & ~~~~~$E'$~~~~~   \\ \hline
		
		$qss-\bar{q}c$ & $i=2$ & $j=7$ & $\Xi^* D^*$          & 3441 & 3444 & ub & 3543 & 3546  \\
		
		$qsc-\bar{q}s$ & $i=2$ & $j=7$ & $\Xi^*_c \bar{K}^*$        & 3466 & 3474 & ub & 3537 & 3545  \\
		
		$ssc-\bar{q}q$ & $i=1$ & $j=7$ & $\Omega_c^* \omega$  & 3558 & 3555 & -3 & 3548 & 3545  \\ \hline\hline
	\end{tabular}
\end{table*}

As is shown in Table~\ref{cc 5/2}, since each structure has only one channel, the channel coupling of one structure is not needed here.
After coupling the $qss-\bar{q}c$ and the $ssc-\bar{q}q$ structure, the energy obtained is still above the threshold.
Besides, after coupling all three channels, a bound state is formed.
The corrected mass of this state is 3527 MeV and the value of RMS of this state is 1.7 fm.
According to the RMS of this state, it tends to be molecular structure and its main composition is $\Xi^*_c \bar{K}^*$.
Therefore, a $J^P=5/2^-$ $ssc\bar{q}q$ pentaquark state is predicted here, whose mass is 3527 MeV.
Although it can decay to some $D$-wave channels, such as $\Xi D$ and $\Xi_c^\prime \bar{K}$, it is still possible to be a resonance, which is worthy of experimental search and research.

\begin{table*}[htb]
	\caption{\label{cc 5/2}The coupled-channel energies of the $ssc\bar{q}q$ pentaquark system with $J^P=\frac{5}{2}^-$ (unit: MeV).}
	\begin{tabular}{c c c c c c} \hline\hline
		
		~~~~Coupled-structure~~~~& ~~~~~~~$E_{th}^{Theo}$ (Channel)~~~~~~~ & ~~~~~~$E_{cc}$~~~~~~ & ~~~~~~~$E_{B}$~~~~~~~ &  ~~~~~~$E_{th}^{Exp}$~~~~~~ & ~~~~~$E'$~~~~~   \\ \hline
		
		$qss-\bar{q}s,~ssc-\bar{q}q$               & 3466 ($\Xi^*_c \bar{K}^*$) & 3469 & ub  & 3537 & 3540  \\
		$qss-\bar{q}c,~qsc-\bar{q}s,~ssc-\bar{q}q$ & 3441 ($\Xi^* D^*$)   & 3430 & -11 & 3537 & 3527  \\ \hline\hline
	\end{tabular}
\end{table*}

\begin{figure}[htb]
	\centering
	\includegraphics[width=8cm]{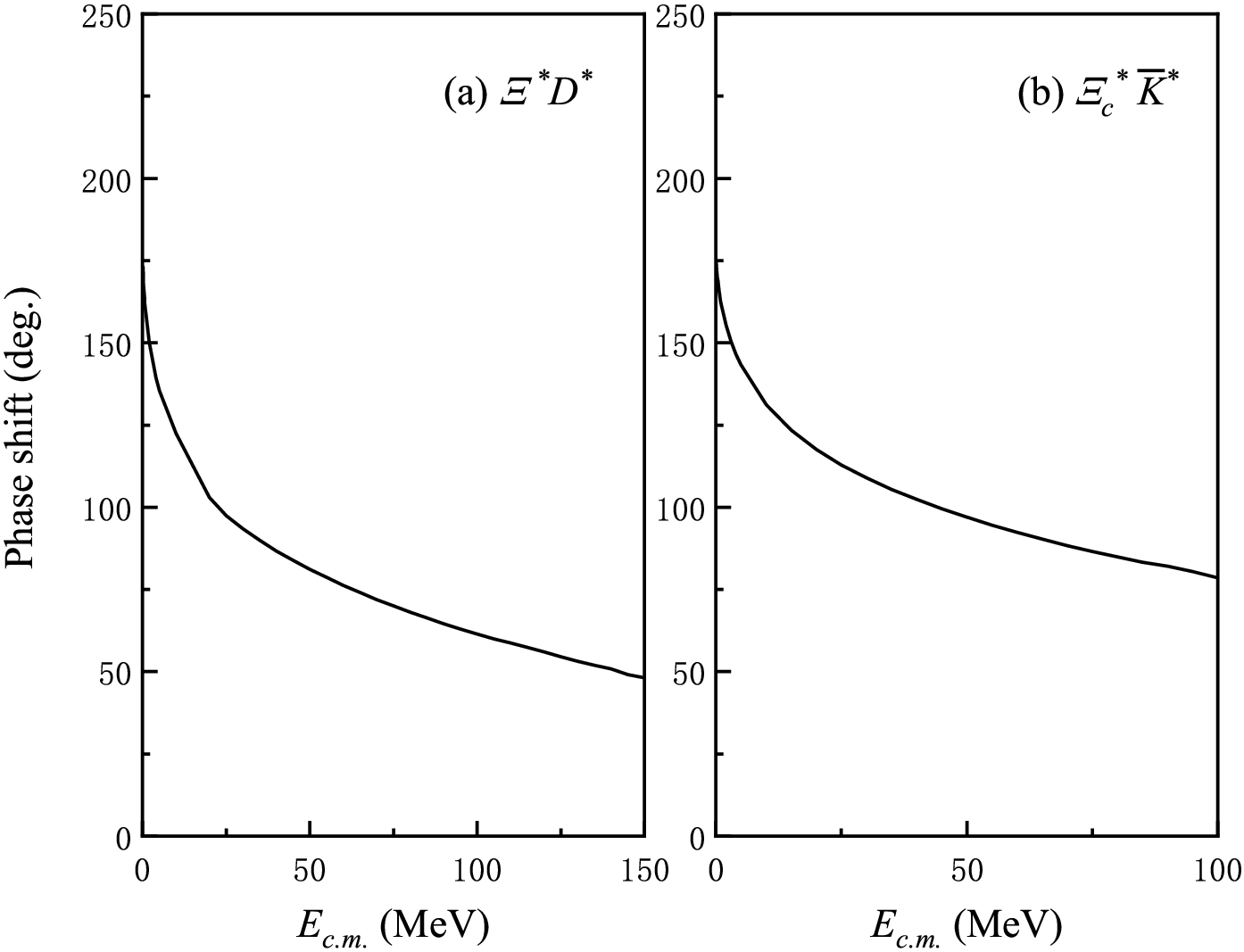}\
	\caption{\label{shift6}  The phase shifts of different open channels with $J^P=\frac{5}{2}^-$.}
\end{figure}

The scattering process is studied to examine whether $\Omega_c^* \omega$ could be a resonance state.
The phase shifts of the $S-$wave open channels  $\Xi^* D^*$ and $\Xi_c^* \bar{K}^*$ are shown in Fig~\ref{shift6}.
However, there is no sharp increase structure of phase shift around the energy of the $\Omega_c^* \omega$ single channel.
This indicates that the $\Omega_c^* \omega$ does not form a resonance state, but rather a scattering state.
In addition, when the incident energy approaches zero, the behavior phase shift also confirms the existence of the bound state.

Considering that there have been a few theoretical works on the newly reported $\Omega_c^0(3185)$ and $\Omega_c^0(3327)$ states, we make a brief review here.
For the $\Omega_c^0(3185)$ state, Refs.~\cite{Yu:2023bxn,Karliner:2023okv,Jakhad:2023mni} interpret it as a three-quark excited state.
The quantum number assignment  $J^P~(nL)$ for the $\Omega_c^0(3185)$ state could be: $J^P=3/2^+~(2S$)~\cite{Yu:2023bxn,Jakhad:2023mni} and $J^P=1/2^+~(2S$)~\cite{Karliner:2023okv}.
On the other hand, the explanation of the $\Omega_c^0(3185)$ as a $\Xi D$ molecular state with $J^P=1/2^-$ can be found in Refs.~\cite{Yan:2023ttx,Xin:2023gkf}.

As for the $\Omega_c^0(3327)$ state, the three-quark excitation explanation can be found in Refs.~\cite{Luo:2023sra,Yu:2023bxn,Wang:2023wii,Karliner:2023okv,Jakhad:2023mni}.
The quantum number assignment for the $\Omega_c^0(3327)$ state could be: $J^P=5/2^+~(1D$)~\cite{Luo:2023sra}, $J^P=3/2^+~(1D$)~\cite{Yu:2023bxn,Jakhad:2023mni}, $J^P=1/2^+, 3/2^+$ or $5/2^+$ ($D-$wave)~\cite{Wang:2023wii} and $J^P=3/2^+~(2S$)~\cite{Karliner:2023okv}.
Refs.~\cite{Feng:2023ixl,Yan:2023ttx,Xin:2023gkf} also support the $\Omega_c^0(3327)$ state to be interpreted as a $\Xi D^*$ molecular state with $J^P=3/2^-$.
The conclusions of the studies on the two newly discovered $\Omega_{c}$ states from different theoretical groups are summarized in Table~\ref{refs}.
According to our calculation, the $\Omega_c^0(3185)$ can be well interpreted as a $\Xi D$ molecular state with $J^P=1/2^-$, whereas $\Omega_c(3327)$ is not found within the multi-quark framework.
Therefore, we propose to explain $\Omega_c(3327)$ from the perspective of the three-quark excitation, and at the same time, the investigation of these states in an unquenched picture could be beneficial.

\begin{table}[htb]
	\caption{\label{refs}The conclusions of the studies on the two newly discovered $\Omega_{c}$ states.}
	\begin{tabular}{c c c c c} \hline\hline
		
		& \multicolumn{2}{c}{$\Omega_c^0(3185)$} & \multicolumn{2}{c}{$\Omega_c^0(3327)$}  \\
		~Ref~                        &three-quark~ & ~molecular~  &~three-quark~ & ~molecular \\ \hline
		
		Ref.~\cite{Luo:2023sra}      &              &              & $\checkmark$ &             \\
		Ref.~\cite{Yu:2023bxn}       & $\checkmark$ &              & $\checkmark$ &             \\
		Ref.~\cite{Wang:2023wii}     &              &              & $\checkmark$ &             \\
		Ref.~\cite{Feng:2023ixl}     &              & $\checkmark$ &              & $\checkmark$ \\
		Ref.~\cite{Karliner:2023okv} & $\checkmark$ &              & $\checkmark$ &             \\
		Ref.~\cite{Yan:2023ttx}      &              & $\checkmark$ &              & $\checkmark$ \\
		Ref.~\cite{Jakhad:2023mni}   & $\checkmark$ &              & $\checkmark$ &             \\
		Ref.~\cite{Xin:2023gkf}      &              & $\checkmark$ &              & $\checkmark$ \\
		This Work                    &              & $\checkmark$ &              &             \\ \hline\hline
	\end{tabular}
\end{table}

In addition to exploring the pentaquark explanation of the two newly discovered $\Omega_{c}$ states, the study of the $ssc\bar{q}q$ system has also led to some other results.
In this work, three states are obtained, including one resonance state and two bound states.
We have summarized the obtained states in Table~\ref{sum}.
Considering that the resonance energy of $\Xi D$ is obtained in the scattering phase shift, the resonance energies obtained in different open channels are not exactly the same.
Therefore, the mass of $\Xi D$ state has a range.
One may notice that there is no value for the decay widths of $\Xi_c^* \bar{K}$ and $\Xi^*_c \bar{K}^*$ in Table~\ref{sum}.
Since $\Xi_c^* \bar{K}$ and $\Xi^*_c \bar{K}^*$ cannot decay to $S$-wave channels, the decay process of the two states to $D$-wave channels will be studied in our future works.
In addition, the decay width to the $D$-wave channels is usually narrow according to our previous research~\cite{Chen:2011zzb}.

\begin{table}[htb]
	\caption{\label{sum} The states obtained in this work.}
	\begin{tabular}{c c c c} \hline\hline
		
		~$J^P$~  & ~~Main Composition~~ & ~~Corrected Mass~~ & ~decay width~   \\ \hline
		
		$1/2^-$  & $\Xi D$              & 3174$\sim$3182 MeV & 42 MeV  \\
		$3/2^-$  & $\Xi_c^* \bar{K}$    & 3138 MeV           &         \\
		$5/2^-$  & $\Xi^*_c \bar{K}^*$  & 3527 MeV           &         \\ \hline\hline
	\end{tabular}
\end{table}

\section{Summary}

In this work, we investigate the excited $\Omega_{c}$ states from the pentaquark perspective.
The $S$-wave pentaquark systems $ssc\bar{q}q$ with $I$ = 0, $J^P = 1/2^-$, $3/2^-$ and $5/2^-$ are studied in the framework of the  QDCSM.
The dynamic bound state calculation is carried out to search for bound states in the $ssc\bar{q}q$ systems.
Both the single-channel and the coupled-channel calculations are performed to explore the effect of the multi-channel coupling.
Meanwhile, the study of the scattering process of the open channels is carried out to confirm possible resonance states.
We also calculate the RMS of cluster spacing to further study the structure of the obtained states.

The numerical results show that a $\Xi D$ resonance state with $J^P=1/2^-$ and two bound states with $J^P=3/2^-$ and $5/2^-$ are obtained.
The mass and the decay width of the $\Xi D$ resonance state is 3174$\sim$3182 MeV and 42 MeV, respectively, which are close to the reported $\Omega_c^0(3185)$.
The RMS of the $\Xi D$ supports the molecular structure of this state. So the recently reported $\Omega_c^0(3185)$ can be explained as the molecular $\Xi D$ state with $J^P=1/2^-$.
It would be very anticipated to see the next experimental steps to determine the spin and parity of it.
A bound molecular state we obtained is $\Xi_c^* \bar{K}$ with $J^P=3/2^-$ and a mass of 3138 MeV, which can be used to interpret the reported $\Omega_c^0(3120)$.
Besides, a new molecular state $\Xi^*_c \bar{K}^*$ with $J^P=5/2^-$ and a mass of 3527 MeV is predicted to exist, which is worth searching in the future.
However, other reported $\Omega_c$ states cannot be well described in the framework of pentaquark systems in present work. The three-quark excitation, or the unquenched picture may be a good explanation, which is worth further exploration.

In addition, the present study shows that the channel coupling effect has to be considered in describing the multi-quark system.
Especially for the possible resonance state, the coupling to the open channels will shift the mass of the resonance state, or even destroy it.
The $\Omega_c \omega$ and $\Omega_c^* \omega$ bound states are obtained in the single-channel calculation.
However, the energies of $\Omega_c \omega$ and $\Omega_c^* \omega$ are elevated by coupling to open channels, leading to the disappearance of these two states.
Based on this, we would like to emphasize the importance of channel coupling effect in studying exotic hadron states.

\acknowledgments{This work is supported partly by the National Science Foundation of China under Contract Nos. 11675080, 11775118, 11535005 and 11865019.}

\setcounter{equation}{0}
\renewcommand\theequation{A\arabic{equation}}

\section*{Appendix A: Resonating group method for bound-state and scattering process}

The resonating group method (RGM)~\cite{RGM1,RGM} and generating coordinates method~\cite{GCM1,GCM2} are used to carry out a dynamical calculation.
The main feature of the RGM for two-cluster systems is that it assumes that two clusters are frozen inside, and only considers the relative motion between the two clusters.
So the conventional ansatz for the two-cluster wave functions is
\begin{equation}
	\psi_{5q} = {\cal A }\left[[\phi_{B}\phi_{M}]^{[\sigma]IS}\otimes\chi(\boldsymbol{R})\right]^{J}, \label{5q}
\end{equation}
where the symbol ${\cal A }$ is the anti-symmetrization operator, and ${\cal A } = 1-P_{14}-P_{24}-P_{34}$. $[\sigma]=[222]$ gives the total color symmetry and all other symbols have their usual meanings.
$\phi_{B}$ and $\phi_{M}$ are the $q^{3}$ and $\bar{q}q$ cluster wave functions, respectively.
From the variational principle, after variation with respect to the relative motion wave function $\chi(\boldsymbol{\mathbf{R}})=\sum_{L}\chi_{L}(\boldsymbol{\mathbf{R}})$, one obtains the RGM equation:
\begin{equation}
	\int H(\boldsymbol{\mathbf{R}},\boldsymbol{\mathbf{R'}})\chi(\boldsymbol{\mathbf{R'}})d\boldsymbol{\mathbf{R'}}=E\int N(\boldsymbol{\mathbf{R}},\boldsymbol{\mathbf{R'}})\chi(\boldsymbol{\mathbf{R'}})d\boldsymbol{\mathbf{R'}},  \label{RGM eq}
\end{equation}
where $H(\boldsymbol{\mathbf{R}},\boldsymbol{\mathbf{R'}})$ and $N(\boldsymbol{\mathbf{R}},\boldsymbol{\mathbf{R'}})$ are Hamiltonian and norm kernels.
By solving the RGM equation, we can get the energies $E$ and the wave functions.
In fact, it is not convenient to work with the RGM expressions.
Then, we expand the relative motion wave function $\chi(\boldsymbol{\mathbf{R}})$ by using a set of gaussians with different centers
\begin{align}
	\chi(\boldsymbol{R}) =& \frac{1}{\sqrt{4 \pi}}\left(\frac{6}{5 \pi b^{2}}\right)^{3 / 4} \sum_{i,L,M} C_{i,L} \nonumber     \\
	&\cdot\int \exp \left[-\frac{3}{5 b^{2}}\left(\boldsymbol{R}-\boldsymbol{S}_{i}\right)^{2}\right] Y_{L,M}\left(\hat{\boldsymbol{S}}_{i}\right) d \Omega_{\boldsymbol{S}_{i}}
\end{align}
where $L$ is the orbital angular momentum between two clusters, and $\boldsymbol {S_{i}}$, $i=1,2,...,n$ are the generator coordinates, which are introduced to expand the relative motion wave function. By including the center of mass motion:
\begin{equation}
	\phi_{C} (\boldsymbol{R}_{C}) = (\frac{5}{\pi b^{2}})^{3/4}e^{-\frac{5\boldsymbol{R}^{2}_{C}}{2b^{2}}},
\end{equation}
the ansatz Eq.~(\ref{5q}) can be rewritten as
\begin{align}
	\psi_{5 q} =& \mathcal{A} \sum_{i,L} C_{i,L} \int \frac{d \Omega_{\boldsymbol{S}_{i}}}{\sqrt{4 \pi}} \prod_{\alpha=1}^{3} \phi_{\alpha}\left(\boldsymbol{S}_{i}\right) \prod_{\beta=4}^{5} \phi_{\beta}\left(-\boldsymbol{S}_{i}\right) \nonumber \\
	& \cdot\left[\left[\chi_{I_{1} S_{1}}\left(B\right) \chi_{I_{2} S_{2}}\left(M\right)\right]^{I S} Y_{LM}\left(\hat{\boldsymbol{S}}_{i}\right)\right]^{J} \nonumber \\
	& \cdot\left[\chi_{c}\left(B\right) \chi_{c}\left(M\right)\right]^{[\sigma]}, \label{5q2}
\end{align}
where $\chi_{I_{1}S_{1}}$ and $\chi_{I_{2}S_{2}}$ are the product of the flavor and spin wave functions, and $\chi_{c}$ is the color wave function.
These will be shown in detail later.
$\phi_{\alpha}(\boldsymbol{S}_{i})$ and $\phi_{\beta}(-\boldsymbol{S}_{i})$ are the single-particle orbital wave functions with different reference centers:
\begin{align}
	\phi_{\alpha}\left(\boldsymbol{S}_{i}\right) & = \left(\frac{1}{\pi b^{2}}\right)^{3 / 4} e^{-\frac{1}{2 b^{2}}\left(r_{\alpha}-\frac{2}{5} \boldsymbol{S}_{i}\right)^{2}}, \nonumber \\
	\phi_{\beta}\left(\boldsymbol{-S}_{i}\right) & = \left(\frac{1}{\pi b^{2}}\right)^{3 / 4} e^{-\frac{1}{2 b^{2}}\left(r_{\beta}+\frac{3}{5} \boldsymbol{S}_{i}\right)^{2}}.
\end{align}
With the reformulated ansatz Eq.~(\ref{5q2}), the RGM Eq.~(\ref{RGM eq}) becomes an algebraic eigenvalue equation:
\begin{equation}
	\sum_{j} C_{j}H_{i,j}= E \sum_{j} C_{j}N_{i,j},
\end{equation}
where $H_{i,j}$ and $N_{i,j}$ are the Hamiltonian matrix elements and overlaps, respectively.
By solving the generalized eigen problem, we can obtain the energy and the corresponding wave functions of the pentaquark systems.

For a scattering problem, the relative wave function is expanded as
\begin{align}
\chi_{L}(\mathbf{R}) & =\sum_{i} C_{i} \frac{\tilde{u}_{L}\left(\boldsymbol{R}, \boldsymbol{S}_{i}\right)}{\boldsymbol{R}} Y_{L,M}(\hat{\boldsymbol{R}}),
\end{align}
with
\begin{align}
\tilde{u}_{L}\left(\boldsymbol{R}, \boldsymbol{S}_{i}\right) & = \left\{\begin{array}{ll}
	\alpha_{i} u_{L}\left(\boldsymbol{R}, \boldsymbol{S}_{i}\right), & \boldsymbol{R} \leq \boldsymbol{R}_{C} \\
	{\left[h_{L}^{-}(\boldsymbol{k}, \boldsymbol{R})-s_{i} h_{L}^{+}(\boldsymbol{k}, \boldsymbol{R})\right] R_{A B},} & \boldsymbol{R} \geq \boldsymbol{R}_{C}
\end{array}\right.
\end{align}
where
\begin{align}
	u_{L}\left(\boldsymbol{R}, \boldsymbol{S}_{i}\right)= & \sqrt{4 \pi}\left(\frac{6}{5 \pi b^{2}}\right)^{3 / 4} \mathbf{R} e^{-\frac{3}{5 b^{2}}\left(\boldsymbol{R}-\boldsymbol{S}_{i}\right)^{2}} \nonumber \\
	& \cdot i^{L} j_{L}\left(-i \frac{6}{5 b^{2}} S_{i}\right).
\end{align}

$h^{\pm}_L$ is the $L$-th spherical Hankel functions, $k$ is the momentum of the relative motion with $k=\sqrt{2 \mu E_{i e}}$, $\mu$ is the reduced mass of two hadrons of the open channel, $E_{i e}$ is the incident energy of the relevant open channels, which can be written as $E_{i e} = E_{total} - E_{th}$ where $E_{total}$ denotes the total energy and $E_{th}$ represents the threshold of open channel.
$R_C$ is a cutoff radius beyond which all the strong interaction can be disregarded.
Besides, $\alpha_i$ and $s_i$ are complex parameters that are determined by the smoothness condition at $R = R_C$ and $C_i$ satisfy $\sum_i C_i = 1$. After performing the variational procedure, a $L$-th partial-wave equationfor the scattering problem can be deduced as
\begin{align}
\sum_j \mathcal{L}_{i j}^L C_j &= \mathcal{M}_i^L(i=0,1, \ldots, n-1),
\end{align}
with
\begin{align}
	 \mathcal{L}_{i j}^L&=\mathcal{K}_{i j}^L-\mathcal{K}_{i 0}^L-\mathcal{K}_{0 j}^L+\mathcal{K}_{00}^L, \nonumber \\
	 \mathcal{M}_i^L&=\mathcal{K}_{00}^L-\mathcal{K}_{i 0}^L,
\end{align}
and
\begin{align}
	\mathcal{K}_{i j}^L= & \left\langle\hat{\phi}_A \hat{\phi}_B \frac{\tilde{u}_L\left(\boldsymbol{R}^{\prime}, \boldsymbol{S}_i\right)}{\boldsymbol{R}^{\prime}} Y_{L,M}\left(\boldsymbol{R}^{\prime}\right)|H-E|\right. \nonumber \\
	& \left.\cdot \mathcal{A}\left[\hat{\phi}_A \hat{\phi}_B \frac{\tilde{u}_L\left(\boldsymbol{R}, \boldsymbol{S}_j\right)}{\boldsymbol{R}} Y_{L,M}(\boldsymbol{R})\right]\right\rangle .
\end{align}
By solving Eq.~(A11), we can obtain the expansion coefficients $C_i$, then the $S-$matrix element $S_L$ and the phase shifts $\delta_L$ are given by
\begin{align}
S_L&=e^{2 i \delta_L}=\sum_{i} C_i s_i.
\end{align}

Resonances are unstable particles usually observed as bell-shaped structures in scattering cross sections of their open channels. For a simple narrow resonance, its fundamental properties correspond to the visible cross-section features: mass $M$ is at the peak position, and decay width $\Gamma$ is the half-width of the bell shape. The cross-section $\sigma_{L}$ and the scattering phase shifts $\delta_{L}$ have relations:
\begin{align}
\sigma_L&=\frac{4 \pi}{k^2}(2 L+1) \sin ^2 \delta_L.
\end{align}
Therefore, resonances can also usually be observed in the scattering phase shift, where the phase
shift of the scattering channels rises through $\frac{\pi}{2}$ at a resonance mass. We can obtain a resonance mass at the position of the phase
shift of $\frac{\pi}{2}$. The decay width is the mass difference between the phase
shift of $\frac{3\pi}{4}$ and $\frac{\pi}{4}$.

\section*{Appendix B: Constructing wave functions}

For the spin wave function, we first construct the spin wave functions of the $q^{3}$ and $\bar{q}q$ clusters with SU(2) algebra, and then the total spin wave function of the pentaquark system is obtained by coupling the spin wave functions of two clusters together.
The spin wave functions of the $q^{3}$ and $\bar{q}q$ clusters are Eq.~(A16) and Eq.~(A17), respectively
\begin{align}
	\chi_{\frac{3}{2}, \frac{3}{2}}^{\sigma}(3) &= \alpha \alpha \alpha, \nonumber\\
	\chi_{\frac{3}{2}, \frac{1}{2}}^{\sigma}(3) & = \alpha \alpha \beta, \nonumber\\
    \chi_{\frac{3}{2}, -\frac{1}{2}}^{\sigma}(3) & = \alpha \beta \beta, \nonumber\\
	\chi_{\frac{3}{2}, -\frac{3}{2}}^{\sigma}(3) &= \beta \beta \beta, \nonumber\\
	\chi_{\frac{1}{2}, \frac{1}{2}}^{\sigma 1}(3) & = \frac{1}{\sqrt{6}}(2 \alpha \alpha \beta-\alpha \beta \alpha-\beta \alpha \alpha),\nonumber\\
	\chi_{\frac{1}{2}, \frac{1}{2}}^{\sigma 2}(3) & = \frac{1}{\sqrt{2}}(\alpha \beta \alpha-\beta \alpha \alpha),\nonumber\\
    \chi_{\frac{1}{2},-\frac{1}{2}}^{\sigma1}(3) & =\frac{1}{\sqrt{6}}(\alpha \beta \beta+\beta \alpha \beta-2 \beta \beta \alpha),   \nonumber\\
	\chi_{\frac{1}{2},-\frac{1}{2}}^{\sigma2}(3) & =  \frac{1}{\sqrt{2}}(\alpha \beta \beta-\beta \alpha \beta).
\end{align}
\begin{align}
	\chi_{1,1}^{\sigma}(2) &  = \alpha  \alpha,   \nonumber \\
	\chi_{1,0}^{\sigma}(2) &  = \frac{1}{\sqrt{2}}(\alpha \beta+\beta \alpha), \nonumber \\
	\chi_{1,-1}^{\sigma}(2) &  = \beta \beta, \nonumber \\
	\chi_{0,0}^{\sigma}(2) & = \frac{1}{\sqrt{2}}(\alpha \beta-\beta \alpha).
\end{align}

For pentaquark system, the total spin quantum number can be 1/2, 3/2 or 5/2.
Considering that the Hamiltonian does not contain an interaction which can distinguish the third component of the spin quantum number, so the wave function of each spin quantum number can be written as follows
\begin{align}
	\chi_{\frac{1}{2}, \frac{1}{2}}^{\sigma 1}(5) = &\chi_{\frac{1}{2}, \frac{1}{2}}^{\sigma}(3) \chi_{0,0}^{\sigma}(2), \nonumber\\
	\chi_{\frac{1}{2}, \frac{1}{2}}^{\sigma 2}(5) = &-\sqrt{\frac{2}{3}} \chi_{\frac{1}{2},-\frac{1}{2}}^{\sigma}(3) \chi_{1,1}^{\sigma}(2)+\sqrt{\frac{1}{3}} \chi_{\frac{1}{2}, \frac{1}{2}}^{\sigma}(3) \chi_{1,0}^{\sigma}(2),   \nonumber\\
	\chi_{\frac{1}{2}, \frac{1}{2}}^{\sigma 3}(5) = & \sqrt{\frac{1}{6}} \chi_{\frac{3}{2},-\frac{1}{2}}^{\sigma}(3) \chi_{1,1}^{\sigma}(2)-\sqrt{\frac{1}{3}} \chi_{\frac{3}{2}, \frac{1}{2}}^{\sigma}(3) \chi_{1,0}^{\sigma}(2) \nonumber \\
	&+\sqrt{\frac{1}{2}} \chi_{\frac{3}{2}, \frac{3}{2}}^{\sigma}(3) \chi_{1,-1}^{\sigma}(2),  \nonumber \\
	\chi_{\frac{3}{2}, \frac{3}{2}}^{\sigma 4}(5) = & \chi_{\frac{1}{2}, \frac{1}{2}}^{\sigma}(3) \chi_{1,1}^{\sigma}(2 ),  \nonumber \\
	\chi_{\frac{3}{2}, \frac{3}{2}}^{\sigma 5}(5) = & \chi_{\frac{3}{2}, \frac{3}{2}}^{\sigma}(3) \chi_{0,0}^{\sigma}(2), \nonumber \\
	\chi_{\frac{3}{2}, \frac{3}{2}}^{\sigma 6}(5) = & \sqrt{\frac{3}{5}} \chi_{\frac{3}{2}, \frac{3}{2}}^{\sigma}(3) \chi_{1,0}^{\sigma}(2)-\sqrt{\frac{2}{5}} \chi_{\frac{3}{2}, \frac{1}{2}}^{\sigma}(3) \chi_{1,1}^{\sigma}(2), \nonumber \\
	\chi_{\frac{5}{2}, \frac{5}{2}}^{\sigma 7}(5) = & \chi_{\frac{3}{2}, \frac{3}{2}}^{\sigma}(3) \chi_{1,1}^{\sigma}(2).
\end{align}

Similar to constructing spin wave functions, we first write down the flavor wave functions of the $q^{3}$ clusters, which are
\begin{align}
	\chi_{0,0}^{f 1}(3) & = \frac{1}{\sqrt{6}}(2 s s c -s c s- c s s), \nonumber \\
	\chi_{0,0}^{f 2}(3) & = \frac{1}{\sqrt{2}}(s c s - c s s), \nonumber \\
	\chi_{0,0}^{f 3}(3) & = \frac{1}{\sqrt{3}}(s s c + s c s + c s s), \nonumber \\
	\chi_{\frac{1}{2}, \frac{1}{2}}^{f 1}(3) & = \sqrt{\frac{1}{6}}(uss+sus-2ssu), \nonumber \\
	\chi_{\frac{1}{2}, \frac{1}{2}}^{f 2}(3) & = \sqrt{\frac{1}{2}}(uss-sus), \nonumber \\
	\chi_{\frac{1}{2}, \frac{1}{2}}^{f 3}(3) & = \sqrt{\frac{1}{3}}(uss+sus+ssu), \nonumber \\
	\chi_{\frac{1}{2}, \frac{1}{2}}^{f 4}(3) & = \sqrt{\frac{1}{12}}(2usc+2suc-csu-ucs-cus-scu), \nonumber \\
	\chi_{\frac{1}{2}, \frac{1}{2}}^{f 5}(3) & = \sqrt{\frac{1}{4}}(ucs+scu-csu-cus), \nonumber \\
	\chi_{\frac{1}{2}, \frac{1}{2}}^{f 6}(3) & = \sqrt{\frac{1}{4}}(ucs+cus-csu-scu), \nonumber \\
	\chi_{\frac{1}{2}, \frac{1}{2}}^{f 7}(3) & = \sqrt{\frac{1}{12}}(2usc-2suc+csu+ucs-cus-scu), \nonumber \\
	\chi_{\frac{1}{2}, \frac{1}{2}}^{f 8}(3) & = \sqrt{\frac{1}{6}}(usc+suc+csu+ucs+cus+scu), \nonumber \\
	\chi_{\frac{1}{2}, -\frac{1}{2}}^{f 1}(3) & = \sqrt{\frac{1}{6}}(dss+sds-2ssd), \nonumber \\
	\chi_{\frac{1}{2}, -\frac{1}{2}}^{f 2}(3) & = \sqrt{\frac{1}{2}}(dss-sds), \nonumber \\
	\chi_{\frac{1}{2}, -\frac{1}{2}}^{f 3}(3) & = \sqrt{\frac{1}{3}}(dss+sds+ssd), \nonumber
\end{align}
\begin{align}
	\chi_{\frac{1}{2}, -\frac{1}{2}}^{f 4}(3) & = \sqrt{\frac{1}{12}}(2dsc+2sdc-csd-dcs-cds-scd), \nonumber \\
	\chi_{\frac{1}{2}, -\frac{1}{2}}^{f 5}(3) & = \sqrt{\frac{1}{4}}(dcs+scd-csd-cds), \nonumber \\
	\chi_{\frac{1}{2}, -\frac{1}{2}}^{f 6}(3) & = \sqrt{\frac{1}{4}}(dcs+cds-csd-scd), \nonumber \\
	\chi_{\frac{1}{2}, -\frac{1}{2}}^{f 7}(3) & = \sqrt{\frac{1}{12}}(2dsc-2sdc+csd+dcs-cds-scd), \nonumber \\
	\chi_{\frac{1}{2}, -\frac{1}{2}}^{f 8}(3) & = \sqrt{\frac{1}{6}}(dsc+sdc+csd+dcs+cds+scd). \nonumber \\
\end{align}
Here, both the light and heavy quarks are considered as identical particles with the SU(4)
extension. Then, the flavor wave functions of $\bar{q}q$ clusters are
\begin{align}
	\chi_{1,1}^{f}(2) & = \bar{d} u, \nonumber \\
	\chi_{1,0}^{f}(2) & = \sqrt{\frac{1}{2}}(\bar{d} d-\bar{u} u), \nonumber \\
	\chi_{1,-1}^{f}(2) & = -\bar{u} d, \nonumber  \\
	\chi_{0,0}^{f}(2) & = \sqrt{\frac{1}{2}}(\bar{d} d+\bar{u} u), \nonumber \\
	\chi_{\frac{1}{2}, \frac{1}{2}}^{f}(2) & = \bar{d} s, \nonumber \\
	\chi_{\frac{1}{2},-\frac{1}{2}}^{f}(2) & = -\bar{u} s, \nonumber \\
	\chi_{\frac{1}{2}, \frac{1}{2}}^{f}(2) & = \bar{d} c, \nonumber \\
	\chi_{\frac{1}{2},-\frac{1}{2}}^{f}(2) & = -\bar{u} c.
\end{align}

As for the flavor degree of freedom, the isospin $I$ of pentaquark systems we investigated in this work is  $I=0$.
The flavor wave functions of pentaquark systems can be expressed as
\begin{align}
	\chi_{0,0}^{f 1}(5) & = \sqrt{\frac{1}{2}} \chi_{\frac{1}{2}, \frac{1}{2}}^{f }(3) \chi_{\frac{1}{2},-\frac{1}{2}}^{f}(2)-\sqrt{\frac{1}{2}}
	\chi_{\frac{1}{2},-\frac{1}{2}}^{f }(3) \chi_{\frac{1}{2}, \frac{1}{2}}^{f}(2),  \nonumber \\
	\chi_{0,0}^{f 2}(5) & = \chi_{0, 0}(3) \chi_{0,0}^{f}(2).
\end{align}

For the color-singlet channel (two clusters are color-singlet), the color wave function can be obtained by $1 \otimes 1$:
\begin{align}
	\chi^{c} =& \frac{1}{\sqrt{6}}(r g b-r b g+g b r-g r b+b r g-b g r)\nonumber \\
	&\cdot\frac{1}{\sqrt{3}}(\bar{r}r+\bar{g}g+\bar{b}b).
\end{align}
Finally, we can acquire the total wave functions by combining the wave functions of the orbital, spin, flavor and color parts together according to the quantum numbers of the pentaquark systems.

\setcounter{equation}{0}
\renewcommand\theequation{B\arabic{equation}}

\end{document}